\documentclass[iop,apj]{emulateapj}

\newcommand\kms{km~s$^{-1}$}
\newcommand\msun{$M_\odot$}

\newcommand\lsun{$L_\odot$}
\newcommand\vsun{$v_\odot$}
\newcommand\vlsr{$v_{LSR}$}

\def\ntot{59}
\def\nnew{5}
\def\nmh{17}

\def\nspring{40}
\def\nfall{19}
\def\be{\begin{equation}}
\def\ee{\end{equation}}
\def\a40{$\alpha$.40}
\def\arcmin{$^{\prime}$}

\def\dg{$^{\circ}$}

\def\ngalfa{11}
\def\fmedian{1.34}
\def\hsizemedian{10}
\def\wmedian{23}

\usepackage{natbib}
\shorttitle{ALFALFA UCHVCs: Local Group Galaxies?}
\shortauthors{E.\ A.\ K.\ Adams et al.}

\begin{document}
\title{A Catalog of Ultra-compact High Velocity Clouds from the ALFALFA Survey: Local Group Galaxy Candidates?}
 
\author {Elizabeth A.\ K.\ Adams\altaffilmark{1}, Riccardo Giovanelli\altaffilmark{1}, Martha P.\ Haynes\altaffilmark{1}}

\altaffiltext{1}{Center for Radiophysics and Space Research, Space Sciences Building,
Cornell University, Ithaca, NY 14853. {\it e--mail:}  betsey@astro.cornell.edu, riccardo@astro.cornell.edu,
haynes@astro.cornell.edu}


\begin{abstract}
We present a catalog of \ntot\ ultra-compact high velocity clouds (UCHVCs) 
extracted from the 40\%-complete ALFALFA HI-line survey.
The ALFALFA UCHVCs have median flux densities of \fmedian\ Jy \kms, 
median angular diameters of \hsizemedian \arcmin, and median velocity widths of \wmedian\ \kms. 
We show that the full UCHVC population
cannot easily be associated with known populations of high velocity clouds.
Of the \ntot\ clouds presented here, only \ngalfa\ are also present in the 
compact cloud catalog extracted from the commensal  GALFA-HI survey,
demonstrating the utility of this separate dataset and analysis.
Based on their sky distribution and observed properties, we infer that the ALFALFA UCHVCs are consistent
with the 
hypothesis that they may  be very low mass galaxies within the Local Volume.
In that case, most of their baryons would be in the form of gas, and because
 of their low stellar content, they remain unidentified by extant optical surveys.
At distances of $\sim$1 Mpc, the UCHVCs have neutral hydrogen (HI) masses of $\sim 10^5 - 10^6$ \msun, 
HI diameters of $\sim 2-3$ kpc, and indicative dynamical masses within the HI extent of $\sim 10^7 - 10^8$ \msun,
similar to the Local Group ultra-faint dwarf Leo T. 
The recent ALFALFA discovery of the star-forming, metal-poor, low mass galaxy Leo P 
demonstrates that this hypothesis is true in at least one case.
In the case of the individual UCHVCs presented here, 
confirmation of their extragalactic nature will  require 
further work, such as the identification of an optical counterpart to
constrain their distance.

\end{abstract}



\section{Introduction} \label{sec:intro}

One of the well-known problems in the study of galaxies is the paucity of
observed low mass galaxies compared to the numbers of them predicted by dark matter simulations.
In the context of the Local Group (LG), this is known as the ``missing satellites''
problem \citep{1999ApJ...522...82K,1999ApJ...524L..19M}.
This discrepancy between simulations and observations is also seen in
the difference between the slope predicted for the low mass end of the dark 
matter halo mass function and the observed slopes of the luminosity function \citep{2005ApJ...631..208B},
neutral hydrogen (HI) mass function \citep{2010ApJ...723.1359M},
and velocity width function \citep{2011ApJ...739...38P}.

During the last decade, much progress has been made in understanding these discrepancies.
The general mismatch between simulations and observations is widely understood to be the result of astrophysical processes impacting the observable baryons.
While simulations are improving at including baryonic physics, many of the relevant processes occur on subgrid scales, leaving many details and specifics as active areas of research.
However, the gross effects of baryon physics are understood.
\cite{2010AdAst2010E..87H} show that simply including the effects of reionization in simulations roughly accounts for the majority of the discrepancy, with baryon content dropping drastically below a critical 
dark matter halo mass of $\sim 10^{10}$ \msun,
near the threshold where galaxy counts are observed to drop dramatically.
The true situation is more complicated; star formation feedback processes are more efficient in massive galaxies but more effective in low mass galaxies so that the baryon content is most depressed at the high and low mass ends of the mass spectrum \citep{2010AdAst2010E..87H,2010MNRAS.404.1111G,2011ApJ...743...45E,2012MNRAS.425.2610R,2012ApJ...759..138P}.

In this context, we distinguish a galaxy from a dark matter halo by the presence of observable baryons.
While the general mismatch between predicted dark matter halos
and visible galaxies is understood, the specifics are not well known.
Is there a minimum galaxy mass that can form? 
Are there galaxies with a single stellar population?
How does star formation proceed in the lowest mass systems?
Which processes are dominant in the baryon loss from the lowest mass systems?
One way to answer these questions is to observe the lowest mass galaxies
that are most impacted by these issues.

The advent of wide-field optical surveys 
increased the number of known Milky Way (MW) satellites
with the discovery of the ultra-faint dwarf galaxies (UFDs).
The UFDs have luminosities from $10^2-10^5$ \lsun, half-light radii from 20-350 pc and $M/L$ ratios of 100 to over 1000, 
total masses within the baryon extent of $10^6-10^7$ \msun, generally old stellar populations, and are located at distances of tens to  a few hundred kpc from the MW
\citep{2008ApJ...684.1075M,2007ApJ...670..313S}.
The name ultra-faint is well earned -- the total luminosities of these objects are comparable to those of globular clusters, 
but they are clearly galaxies as their kinematics indicate they are dark matter dominated \citep{2007ApJ...670..313S}.
The discovery of UFDs is exciting and opens many possibilities into addressing the fundamental 
questions of how marginal galaxies form; however there is one problem --
 nearly all the UFDs are located within the virial radius of the MW.
\cite{BR11} predict based on simulations that the vast majority of UFDs have been modified by tides;
this is supported by observational evidence of tidal disruption
\citep{2007ApJ...670..313S,2010AJ....140..138M,2012ApJ...756...79S}.
This makes it nearly impossible to determine which of the UFD properties, such as sizes and kinematics,
 are primeval
 and which are result of environmental influence
from interaction with the MW. 
\cite{BR11} do predict the existence of $\sim$100 fossil galaxies with luminosities less than $10^6$ \lsun\ 
that have remained isolated from 
the Milky Way at distances of 400 kpc to 1 Mpc.

One UFD is 
 of particular note. Leo T lies at distance of 420 kpc, safely outside the virial radius of the MW and was, until recently, the only gas-rich UFD discovered.
Leo T is a star-forming galaxy with a HI mass of 2.8 $\times$ $10^5$ \msun, an HI diameter of 600 pc, 
an indicative dynamical mass within the HI extent of $\sim$3.3$\times$ $10^6$ \msun, a total-mass-to-light ratio within the HI extent of 56, 
and a stellar mass of $\sim$1.2 $\times$ $10^5$ \msun\ \citep{2008MNRAS.384..535R}.
Given its gas content and distance, Leo T likely represents an unperturbed UFD, allowing environmental effects to be disentangled from the evolution of the lowest-mass galaxies.
Indeed, \cite{2012MNRAS.425..231R} argue that Leo T is on its first infall to the Milky Way.
Leo T is on the edge of detectability for SDSS; were it located further away, its stellar population would not have been detected \citep{2010AdAst2010E...8K}.
UFDs with properties similar to Leo T but located further from the MW
or with fainter stellar populations would have been overlooked in the automated searches of SDSS.
However, the HI content of Leo T would be detectable in a sensitive, wide area HI survey, raising the possibility that
more isolated, gas-rich UFDs await discovery.

Exploiting the huge collecting area of the Arecibo 305m 
telescope\footnote{The Arecibo Observatory is operated by SRI 
International under a cooperative agreement with the National Science Foundation (AST-1100968), 
and in alliance with Ana G. M\'{e}ndez-Universidad Metropolitana, 
and the Universities Space Research Association.} 
and the mapping capability of its 7 beam receiver (ALFA),
the Arecibo Legacy Fast ALFA (ALFALFA) HI line survey is the first blind HI survey capable of addressing this issue in a robust way.
Surveying over 7000 square degrees of sky, ALFALFA has the sensitivity to detect $10^5$ \msun\  of HI with 
a linewidth of 20 \kms\  at 1 Mpc.
In fact, the recent discovery of Leo P from ALFALFA survey data shows that galaxies similar to Leo T in the Local Volume may be identified via their 21cm line emission. 
\citep{LeoP_HI,LeoP_optical,LeoP_metallicity}.
Leo P was discovered during the normal course of identifying HI detections within the ALFALFA survey
when it was noticed that one 
ultra-compact high velocity cloud (UCHVC) could be associated with an
 irregular, lumpy light distribution in the SDSS images \citep{LeoP_HI}.
Follow-up optical observations resolved a stellar population and a single HII region, confirming that the
UCHVC is in fact a low mass galaxy, Leo P \citep{LeoP_optical}.
We stress that Leo P was confirmed to be a galaxy because its young, blue stellar population was barely visible 
in the SDSS images; without recent star formation, the underlying older population of
Leo P would not have been visible at all in the SDSS images.
Leo P was discovered by its HI signature, and 
its existence strongly argues that other very low mass and (nearly) starless 
objects are included among the ALFALFA UCHVCs.

We 
 \citep[hereafter G10]{2010ApJ...708L..22G} originally discussed a set of ultra-compact high velocity clouds (UCHVCs) that were consistent with being gas-bearing 
low mass dark matter halos at $\sim$1 Mpc; we referred to this interpretation of the
UCHVCs as the minihalo hypothesis. 
In this paper, we expand on this work and
present a catalog of UCHVCs for the current 40\% ALFALFA data release, termed \a40 \citep{2011AJ....142..170H}.
We offer further detail on the minihalo hypothesis for this class of objects,
drawing special attention to the properties of Leo T and Leo P.
We note that the idea that LG dwellers could be identified by their HI content was first proposed by 
\citet{1999A&A...341..437B} and \citet{1999ApJ...514..818B}.
The UCHVCs presented here overcome objections raised against the initial sample of clouds proposed to
represent gas-rich galaxies in the LG.

In Section \ref{sec:data} we discuss the \a40 data and selection of UCHVCs.
In Section \ref{sec:catalog} we present the UCHVC catalog and overview the observed properties of the UCHVCs.
In Section \ref{sec:other} we examine the UCHVC population in the context of 
the known high velocity cloud (HVC) populations,
and in Section \ref{sec:mhc} we present evidence supporting the LG origin and minihalo hypothesis for the UCHVCs.
In Section \ref{sec:conclusion}, we summarize our findings.


\section{Data} \label{sec:data}
\begin{figure*}[th]
\begin{center}
\includegraphics[keepaspectratio,width=\linewidth,trim=0cm 0cm 2cm 0cm]{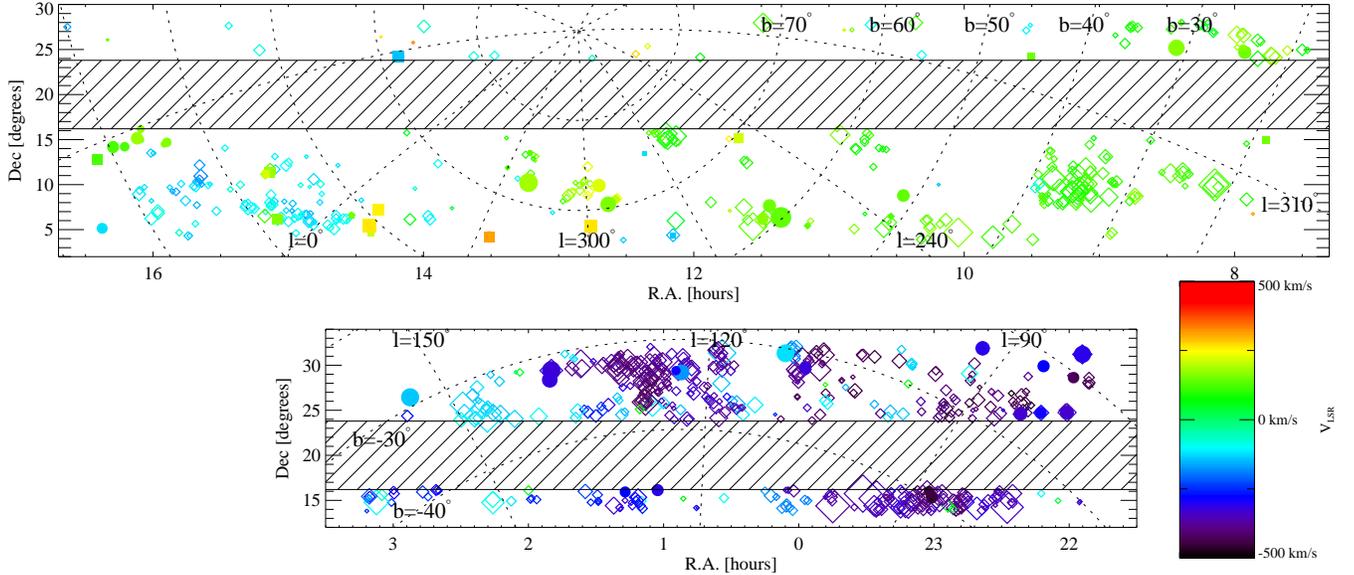}
\caption{UCHVCs (filled circles) plotted in R.A.-Dec. coordinates; 
gray scale (color in the online version) 
corresponds to the velocity of the cloud. 
The solid squares are the most-isolated subsample of UCHVCs (see Section \ref{sec:iso}).
The open diamonds are the \a40 HVCs shown for reference.
The size of the symbols is proportional to the angular sizes of the HVCs in all cases but not to scale.
The top panel is the spring R.A. region; the bottom panel the fall R.A. region. 
The hashed region corresponds to declination ranges not covered by \a40.
The fall sky shows prevalent HVC structure while the spring sky is relatively clear of HVCs.
}
\label{fig:radec}
\end{center}
\end{figure*}

The sources presented here are found within the footprint of the \a40 release of the ALFALFA survey \citep{2011AJ....142..170H}
but correspond to a separate analysis of the same spectral data cubes.
We briefly describe the ALFALFA survey below, with an emphasis on its relevance to UCHVCs, 
followed by a description of how the UCHVCs are identified and measured.
The ALFALFA sky is divided into two regions, termed the ``spring'' and ``fall''  
as a result of our nighttime observing in the Northern Hemisphere.
The ``spring'' ALFALFA sky covers a range of $7.5h-16.5h$ in RA; the ``fall'' sky is $22h-3h$ in RA.
The \a40 footprint covers approximately 2800 square degrees and
includes the declination ranges 4\dg-16\dg\ and 24\dg-28\dg\  in the spring, and 14\dg-16\dg\ and 24\dg-32\dg\ in the fall.
We note here that Leo P is located at +18\dg\ and is not in the \a40 footprint, and hence is not included in  the UCHVC sample.
The footprint of the \a40 survey can be seen in Figure \ref{fig:radec};
the top panel is the spring sky and the bottom panel is the fall sky.
The relative sizes of the panels indicate the different RA coverage of the separate survey areas.
The open diamonds in the figure show the general HVC population of the \a40 survey
and the filled symbols are the UCHVCs of this work with the gray scale (color in the online version)
indicating the velocities of the clouds.
The fall sky 
shows a prevalence of HVCs;
in comparison, the spring sky is relatively clean, making this a better location to look
for low mass gas-bearing dark matter halos.

\subsection{The ALFALFA Survey}

ALFALFA is an extragalactic spectral line survey 
making use of the Arecibo 305m telescope.
The survey maps 7000 square degrees of sky in the HI 21cm line, covering the
spectral range between 1335 and 1435 MHz (roughly -2500 \kms\ to 17500 \kms\ for
the HI line), with a spectral resolution of 25 kHz, or $\sim 5.5$ \kms\ (at $z=0$).
ALFALFA is designed to outperform previous blind HI surveys.
With an angular resolution of $\sim 3'.5$, ALFALFA can resolve structures 1/4
the angular size possible with the HI Parkes All Sky Survey \citep[HIPASS; 
][]{2004MNRAS.350.1195M}
 and 1/9 that possible with the Leiden Dwingeloo Survey \citep[LDS; 
][]{1997agnh.book.....H}. Its flux density
sensitivity is nearly one order of magnitude higher than that of HIPASS 
and more than two orders of magnitude better than that of the LDS. ALFALFA can detect
a $\sim 5\times 10^4$ \msun\ cloud of 20 \kms\ linewidth at a distance of 1 Mpc.
A full description of the observational mode of ALFALFA is given in \cite{2007AJ....133.2569G},
while the definition and goals of the survey are described in \cite{2005AJ....130.2598G}.
Only a summary of the observational details is given here.

ALFALFA surveys the sky using a
seven--feed multi-beam receiver in ``drift'' mode: the telescope is normally parked along the local
meridian and 14 tracks (7 feeds, 2 polarizations each) of spectral data of 4096 channels --
each acquired continuously and recorded at a 1 Hz rate as the sky drifts by. All regions of
the sky are visited twice with  the two visits typically a few months apart
in time. Upon completion of data taking of a region of the sky,
data cubes of 2\dg.4 $\times$ 2\dg.4 in spatial coordinates are produced and
sampled over a regular grid of 1\arcmin\ spacing in R.A. and Dec. 
After Hanning smoothing to 11 \kms\ resolution, the rms noise per channel
of the data is typically
2 to 2.5 mJy per beam. In general, sources are extracted from the data cubes through a 2--step process. An
automated signal identification algorithm is first run over each data cube, producing a
preliminary source catalog \citep{2007AJ....133.2087S}. Then each source in the catalog is visually
inspected and remeasured. The measurement tool fits ellipses to contours of constant flux density
level and delivers a source position, given by the center of the ellipse encircling
half of the total flux density of the source, source sizes (as the major and minor axes of
said ellipse), flux density, velocity and linewidth.

\subsection{Source Identification}
As mentioned above, the standard source identification and measurement in ALFALFA uses the algorithm developed by 
\cite{2007AJ....133.2087S} to identify sources and is then followed by measurement of the source by hand. 
Briefly, the identification algorithm is a one-dimensional matched filtering scheme. 
The spectrum in each pixel of an ALFALFA grid is matched to a series of Hermite polynomial templates. 
The detection of a galaxy requires the detection of spectra of similar velocity widths with a high significance in 
5 or more contiguous pixels.

\begin{figure*}[t]
\begin{center}
\includegraphics[keepaspectratio,width=0.6\linewidth]{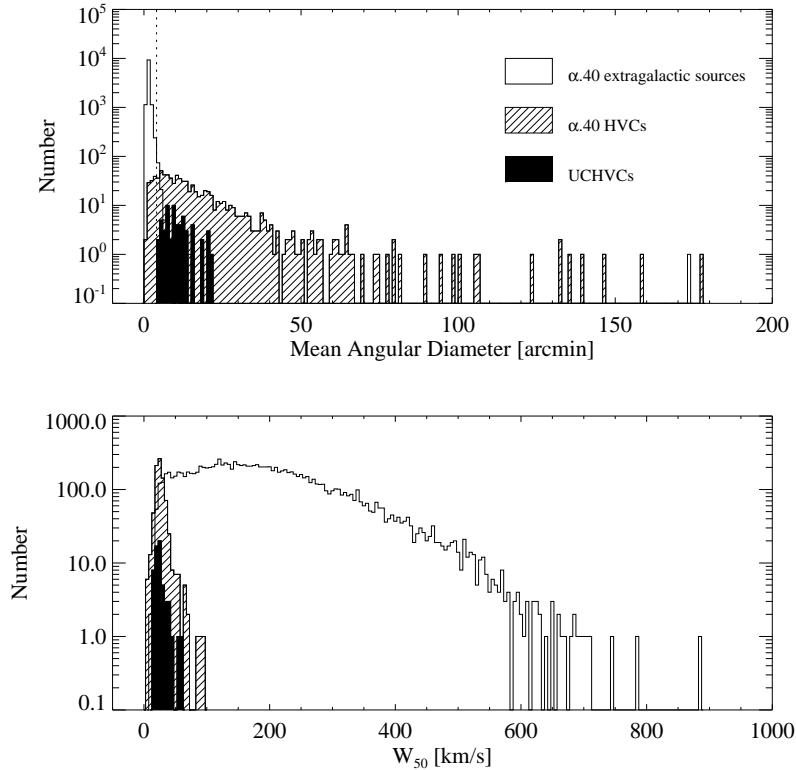}
\caption{The distribution of HI angular diameters and full width half maximum of the HI line ($W_{50}$)
 for the UCHVCs (filled histograms), \a40 sources classified as 
reliable  extragalactic detections (unfilled histograms),
and \a40 HVCs (hashed histograms). The UCHVCs and HVCs occupy a small range of narrow velocity widths. 
The UCHVCs are spatially large compared to the extragalactic detections but generally
small compared to the \a40 HVCs.}
\label{fig:compare1_9}
\end{center}
\end{figure*}

In comparison with the extragalactic sources identified in the \a40 catalog, 
the UCHVCs are typically spatially extended and have narrow velocity widths. 
This is illustrated in Figure \ref{fig:compare1_9} where the distribution of HI angular diameters 
and velocity widths are plotted
for the \a40 extragalactic sources, \a40 HVCs and the UCHVCs of this work.
The UCHVCs are spatially extended compared to the extragalactic sources but generally small compared 
to the full HVC population of the \a40 survey.
The minimum velocity width used in the templates of the \cite{2007AJ....133.2087S} identification algorithm is 30 \kms, the typical \emph{maximum} width of the UCHVCs. 
For this reason, a special source identification algorithm was developed for the UCHVCs in addition to the standard ALFALFA pipeline. 
This method is based on the philosophy of \cite{2007AJ....133.2087S},
but with
three main differences: a limited velocity range, three-dimensional matched filtering, and the use of gaussian templates. 
Only a limited velocity range of the ALFALFA data set, -500 \textless\ \vsun\ \textless\ 1000 \kms, is selected as this is the expected velocity range for objects within the Local Volume.  
Because only a limited velocity range is examined, it is reasonable to perform a full three-dimensional matched filtering, 
matching both the spectrum of the source and the spatial position and size simultaneously. 
Gaussian templates are used to describe both the spatial extent and the velocity profile of the UCHVCs.
The templates range from a spatial full width half maximum  (FWHM) size of 4\arcmin\ to 12\arcmin\  in steps of 2\arcmin\  
and the spectral line FWHM ranges from  10 \kms\ to 40 \kms\  in steps of 6 \kms.
The lower bound of the spatial templates is set by the beam size of Arecibo. 
The upper size bound is near the median size value of the UCHVCs and represents our emphasis on detecting ultra-compact clouds.
UCHVCs can be larger in size than 12\arcmin\  and the matched filtering of the 12\arcmin\ template to
 a UCHVC with HI diameter greater than 12\arcmin\ is robust.
A velocity FWHM of 10 \kms\  represents the narrowest source that can be spectroscopically resolved in the ALFALFA data.
The warm neutral medium is thought to be the dominant phase of the ISM in minihalos \citep[e.g.][]{2002ApJS..143..419S};
 for a reasonable range of temperatures ($6000-10000$ K) for the warm neutral medium
in the UCHVCs, thermal broadening results in linewidths of $\sim16-21$ \kms.
Thus for a cloud of 40 \kms\ linewidth, we would expect the large scale motion to be $\sim$34 \kms\ for
the warmest clouds, after subtracting the thermal broadening contribution in quadrature.
For a typical size of 10\arcmin\ at an indicative distance of 1 Mpc, the dynamical mass 
based on this unbroadened linewidth is $\sim 10^8$ \msun.
This is a reasonable upper limit to the dynamical mass we may expect to be traced out 
for a more massive dark matter halo
of $\lesssim 10^{10}$ \msun, and matches the dynamical mass traced by the baryon extent of
the presumably more massive SHIELD galaxies \citep{2011ApJ...739L..22C}.

\begin{figure*}
\begin{center}
\includegraphics[keepaspectratio,trim=1cm 0cm 0cm 0cm,width=0.6\linewidth]{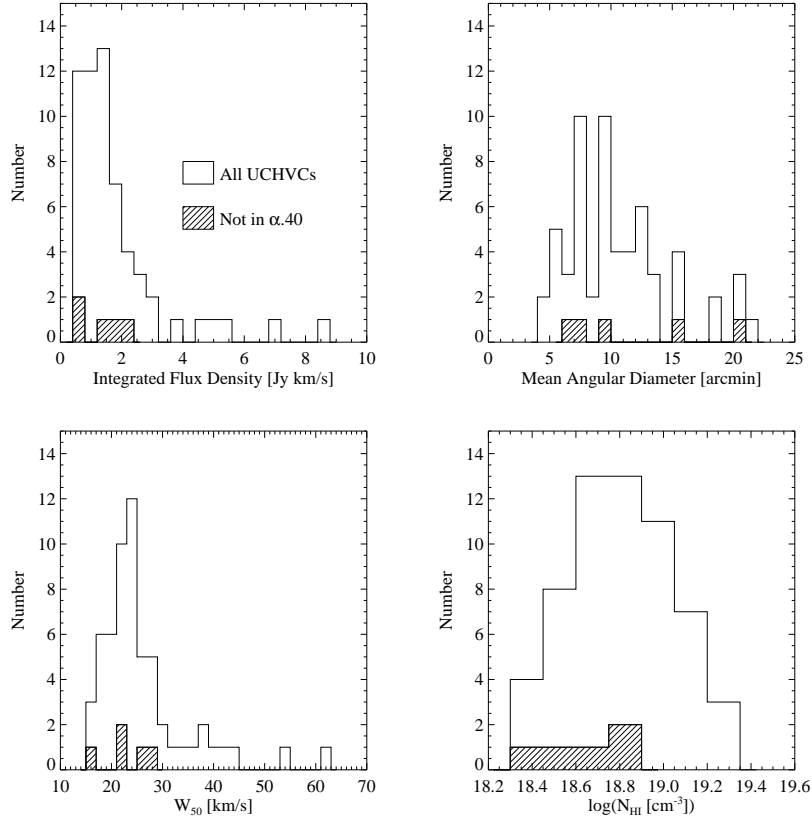}
\caption{
The measured properties for the full sample of UCHVCs (unfilled histograms) compared to the UCHVCs found specifically by 
the separate analysis presented in this work (hashed histograms). Generally, the new UCHVCs have narrow velocity widths and low fluxes.
They also have low $\bar N_{HI}$ values. 
}
\label{fig:compare_uchvcs}
\end{center}
\end{figure*}

Many of the UCHVCs are missed by the standard identification algorithm.
Since the ALFALFA pipeline also involves visual inspection of the dataset, 
most of these sources are identified by eye and 
included in the \a40 catalog as HVC detections.
The specialized UCHVC identification algorithm does find sources that are missed by the standard ALFALFA pipeline;
of the \ntot\  UCHVCs identified here (listed in Table \ref{tab:uchvc}),  \nnew\    sources are not included in the \a40 catalog.
Three of these are in the spring sky and two in the fall sky.
Figure \ref{fig:compare_uchvcs} shows the measured properties of all the UCHVCs compared to the \nnew\ sources not included in the \a40 catalog.
The additional sources tend to have low integrated flux densities and narrow linewidths ($W_{50}$). 
While they have a range of HI diameters, they are not the most compact clouds.
Most strikingly, the UCHVCs not included in the \a40 catalog
 are the sources with the lowest average column densities, 
suggesting that these sources are the tip of the iceberg for further clouds to be detected.

\subsection{Criteria for UCHVC Identification}
To be included as a UCHVC in the catalog, a source must have $|v_{LSR}|$ \textgreater\ 120 \kms, have a HI major axis less than 30\arcmin\ in size, and have a S/N $\geq$ 8 to ensure reliability. 
The \vlsr\ limit is imposed  to focus on a class of clouds that are well separated from Galactic emission and
that could trace dark matter halos within the LG. 
Some dark matter halos would be expected to have $|$\vlsr$|$ \textless\ 120 \kms\ (Leo T, for example)
but disentangling their emission from Galactic hydrogen is challenging and left to future work.
The 30\arcmin\ size limit corresponds to a physical size of 2 kpc at a distance of 250 kpc.
The distance of 250 kpc is a reasonable minimum distance for an unperturbed object at the edge of the MW;
\cite{2009ApJ...696..385G} find that LG dwarf galaxies with neutral gas content 
270 kpc away from either the MW or M31\footnote{The Magellanic Clouds do have a substantial neutral gas content and are much closer to the MW than 250 kpc. However, they are more massive than the general population of dwarfs in the LG and are actively losing their HI via interactions with the MW.}.
We would not expect to detect low mass galaxies with large gas reservoirs nearer to the MW due to interaction with the hot Galactic corona \citep{2006ApJ...639..590F}.
Note that Leo T has an HI diameter of 0.6 kpc, and Leo P has an HI diameter of 1.2 kpc. 
The models of \cite{2002ApJS..143..419S} for gas in dark matter minihalos predict HI diameters up to 3 kpc with diameters less than 2 kpc a more common outcome.
It should be noted that most UCHVCs are smaller than this criterion, with only six clouds 
having average HI diameters larger than 16\arcmin\ (see Figure \ref{fig:compare_uchvcs}).
The S/N limit of 8 ensures reliability. 
This limit is higher than the general reliability limit of the ALFALFA survey data due to the different nature
of the UCHVCs, including the strong potential for radio frequency interference (RFI)
 to masquerade as narrow-line sources.
Confirmation observations of low S/N sources are ongoing, and in future work we will examine the reliability and completeness of the ALFALFA UCHVC catalog.

\subsection{Isolation}\label{sec:iso}
Given the abundance of HVC structures in the sky, 
the most important criterion for determining if a cloud is a good minihalo candidate is its isolation.
Most of the known HVC structure is associated with Galactic processes, 
including accretion onto the Milky Way;
when considering clouds that could represent gas associated with dark matter halos,
we wish to find objects distinct from existing HVC structures.
%
In order to be considered a UCHVC, 
visual inspection must ensure that the cloud does not appear to be associated with a larger HI structure.


\begin{deluxetable*}
{lcccccccrrl} 
\tablewidth{0pt}
\tabletypesize{\scriptsize}
\tablecaption{ALFALFA UCHVCs in the $\alpha$40 Survey \label{tab:uchvc}}
\tablehead{
\colhead{Source}  & AGC    & \colhead{R.A.+Dec.} &
\colhead{cz$_\odot$ $V_{lsr}$ $V_{gsr}$ $V_{LG}$} & \colhead{$W_{50}(\epsilon_w)$}  &
\colhead{$a\times b$} & \colhead{$S_{21}$}      & \colhead{S/N} & $N_3$ & $N_{10}$ &\colhead{Notes}    \\
{} & {} & J2000 &  \kms & \kms & \arcmin & Jy \kms &  &  &  &  
}
\startdata
HVC111.65-30.53-124\tablenotemark{a} & 103417 & 000554.3+312014 &  -128 -124   \ \ 55 \ 139 &  21 ( 8) &  27 $\times$ 15 &  2.31  & 12 &  0 &  9 &  g, O \\
HVC123.11-33.67-176 & 102992 & 005206.2+291204 &  -177 -176  \ -19  \ \ \ 61 &  21 ( 3) &  24 $\times$ 10 &  1.28  &  9 &  0 &  5 &  g, O \\
HVC123.74-33.47-289\tablenotemark{c} & 102994 & 005431.6+292402 &  -290 -289 -133  \ -52 &  21 ( 1) & \  6 $\times$ \ 5 &  0.67  & 15 &  0 & 13 &  g, O \\
HVC126.85-46.66-310 & 749141 & 010237.8+160752 &  -308 -310 -186 -112 &  23 ( 6) &  10 $\times$ \ 8 &  0.81  &  9 &  1 & 21 &  g, O \\
HVC131.90-46.50-276\tablenotemark{a,c} & 114574  & 011703.4+155548 &  -273 -276 -160  \ -88 &  27 ( 4) &   10 $\times$ \ 6 &  0.71  &  9 &  1 & 22 &  g, O \\
HVC137.90-31.73-327 & 114116 & 014952.1+292600 &  -325 -327 -199 -124 &   34 ( 8) &  29 $\times$ 16 &  3.93  & 13  & 1 & 9 & g, O \\
HVC138.39-32.71-320 & 114117 & 015031.4+282259 &  -317 -320 -194 -119 &   22 ( 2) &  19 $\times$ 13 &  4.41  & 30  & 1 & 7 & g, O, S12 \\
HVC154.00-29.03-141 & 122836 & 025229.7+262630 &  -135 -141  \ -55  \ \ \ \ 8 &  27 ( 3) &  29 $\times$ 15 &  6.90  & 31 &  0 & 15 &  g, O \\
HVC205.28+18.70+150\tablenotemark{$\bigstar$} & 174540 & 074559.9+145837 &  \ 162 \ 150   \ \ 59 \ \ 42 &  23 ( 4) &  10 $\times$  6 &  2.06  & 28 &  0 &  2 &  g, S12, O \\
HVC196.50+24.42+146 & 174763 & 075527.1+244143 &   \ 156 \ 146  \ \ 88 \ \ 79 &  20 ( 2) &  16 $\times$ 11 &  2.80  & 20 &  3 & 11 &  g, S12, O \\
HVC196.09+24.74+166 & 174764 & 075614.8+250900 &  \ 175 \ 166 \ 110 \ 101 &  24 ( 6) &  10 $\times$ \ 5 &  0.66  &  9 &  3 & 10 &  p, O \\
HVC198.48+31.09+165 & 189054 & 082546.7+251128 &  \ 173 \ 165 \ 104 \ \ 90 &  26 ( 1) &  19 $\times$ 13 &  1.77 & 13 &  0 &  8 &  g, O \\
HVC204.88+44.86+147\tablenotemark{$\bigstar$} & 198511 & 093013.2+241217 &  \ 152 \ 147 \ \ 80 \ \ 53 &  15 ( 1) &  \ 8 $\times$ \ 6 &  0.73 & 14 &  0 &  0 &  g, S12, O \\
HVC234.33+51.28+143 & 208315 & 102701.1+084708 &  \ 148 \ 143 \ \ 29 \ -22 &  20 ( 2) &  15 $\times$ 10 &  4.96  & 35 &  0 & 16 &  g, S12 \\
HVC250.16+57.45+139 & 219214 & 110929.8+052601 &  \ 142 \ 139  \ \ 25 \ -32 &  20 ( 5) & \  7 $\times$ \ 4 &  0.56 & 10 &  0 &  9 &  g, G10, S12 \\
HVC252.98+60.17+142 & 219274 & 112119.6+062132 &  \ 143 \ 142  \ \ 35 \ -22 &  27 ( 5) &  28 $\times$ 15 &  8.55 & 37 &  1 & 10 &  g, S12 \\
HVC253.04+61.98+148 & 219276 & 112624.8+073915 &  \ 149 \ 148 \ \ 47  \ \ -8 &  36 ( 1) &  14 $\times$ 12 &  2.06 & 14 &  1 & 11 &  g \\
HVC255.76+61.49+181 & 219278 & 112855.6+062529 &  \ 182 \ 181 \ \ 77  \ \ 19 &  18 ( 2) &  11 $\times$ \ 6 &  0.90 & 13 &  0 &  7 &  g, S12 \\
HVC256.34+61.37+166\tablenotemark{c} & 219279 & 112928.6+060923 &  \ 167 \ 166 \ \ 61 \ \ \  3 &  24 ( 1) &  12 $\times$ 11 &  1.49 & 14 &  2 & 11 &  g \\
HVC245.26+69.53+217\tablenotemark{$\bigstar$} & 215417 & 114008.1+150644 &  \ 216 \ 217 \ 146 \ \ 97 &  17 ( 4) &  10 $\times$ \ 9 &  0.70  &  9 &  0 &  1 &  g, G10 \\
HVC277.25+65.14-140\tablenotemark{$\bigstar$} & 227977 & 120920.0+042330 &  -142 -140 -234 -294 &  23 ( 1) & \  7 $\times$ \ 4 &  0.46  &  8 &  0 &  1 &  g, G10 \\
HVC274.68+74.70-123\tablenotemark{$\bigstar$} & 226067 & 122154.7+132810 &  -128 -123 -182 -232 &  54 (13) & \  5 $\times$ \ 4 &  0.92  & 11 &  0 &  0 &  p, G10 \\
HVC290.19+70.86+204 & 226165 & 123440.2+082408 &  \ 200 \ 204 \ 135  \ \ 80 &  21 ( 1) &   10 $\times$ \ 6 &  0.90  & 11 &  1 & 15 &  g \\
HVC292.94+70.42+159\tablenotemark{a}  & 229344 & 123758.5+074849 &  \ 154 \ 159 \ \ 89 \ \ 34 &  15 ( 4) &  17 $\times$ 14 &  1.67  & 13 &  0 & 18 &  g \\
HVC295.19+72.63+225 & 226170 & 124204.6+095405 &  \ 220 \ 225 \ 164 \ 112 &  28 ( 7) &  14 $\times$ 12 &  1.17  & 10 &  3 & 16 &  p, G10 \\
HVC298.95+68.17+270\tablenotemark{$\bigstar$} & 227987 & 124529.8+052023 &  \ 265 \ 270 \ 196 \ 139 &  26 ( 1) &  16 $\times$ \ 9 &  5.58  & 44 &  0 &  4 &  g, G10 \\
HVC324.03+75.51+135 & 233763 & 131242.3+133046 & \  127 \ 135 \ 102 \ \ 56 &  29 ( 1) & \  7 $\times$ \ 5 &  0.94  & 18 &  1 & 12 &  g \\
HVC320.95+72.32+185 & 233830 & 131321.5+101257 &  \ 177 \ 185 \ 141 \ \ 92 &  23 ( 9) &  21 $\times$ 16 &  1.70  &  9 &  0 & 15 &  g, G10 \\
HVC330.13+73.07+132 & 233831 & 132241.6+115231 &  \ 124 \ 132 \ 100 \ \ 53 &  16 ( 1) & \  6 $\times$ \ 3 &  0.63  & 11 &  0 & 11 &  g, G10 \\
HVC326.91+65.25+316\tablenotemark{$\bigstar$} & 238713 & 133043.8+041338 & \  308 \ 316 \ 264 \ 210 &  26 ( 4) &  12 $\times$  10 &  1.25  & 11 &  0 &  0 &  p, G10 \\
HVC 28.09+71.86-144\tablenotemark{$\bigstar$} & 249393 & 141058.1+241204 &  -157 -144 -111 -136 &  43 ( 6) &  15 $\times$ \ 9 &  1.12  &  8 &  0 &  0 &  g, O \\
HVC353.41+61.07+257\tablenotemark{$\bigstar$} & 249323 & 141948.6+071115 &  \ 246 \ 257 \ 244 \ 201 &  20 ( 4) &  13 $\times$ \ 9 &  1.34  & 13 &  3 &  4 &  g, G10 \\
HVC351.17+58.56+214\tablenotemark{$\bigstar$,b} & 249282 & 142321.2+043437 &  \ 203 \ 214 \ 196 \ 151 &  40 ( 8) &  \ 7 $\times$ \ 5 &  1.45  & 17 &  0 &  4 &  p, G10, S12 \\
HVC352.45+59.06+263\tablenotemark{$\bigstar$} & 249283 & 142357.7+052340 &  \ 252 \ 263 \ 248 \ 203 &  32 ( 9) &  16 $\times$ 11 &  1.11  &  8 &  3 &  4 &  g, G10 \\
HVC356.81+58.51+148\tablenotemark{$\bigstar$} & 249326 & 143158.8+063520 &  \ 136 \ 148 \ 141 \ 100 &  38 (11) &  \ 6 $\times$ \ 5 &  0.70  &  10 &  0 &  1 &  p \\
HVC  5.58+52.07+163\tablenotemark{$\bigstar$} & 258459 & 150441.3+061259 &  \ 149 \ 163 \ 176 \ 141 &  24 ( 8) &  11 $\times$  10 &  1.33 & 13 &  0 &  4 &  g \\
HVC 13.59+54.52+169\tablenotemark{$\bigstar$} & 258237 & 150723.0+113256 &  \ 155 \ 169 \ 200 \ 170 &  23 ( 3) &  10 $\times$ \ 5 &  1.34  & 17 &  1 &  3 &  g \\
HVC 13.60+54.23+179\tablenotemark{$\bigstar$} & 258241 & 150824.4+112422 &  \ 164 \ 179 \ 210 \ 180 &  17 ( 1) &  15 $\times$ \ 7 &  0.99  & 11 &  1 &  4 &  g \\
HVC 13.63+53.78+222\tablenotemark{$\bigstar$} & 258242 & 151000.6+111127 & \  207 \ 222 \ 253 \ 224 &  21 ( 2) &  \ 9 $\times$ \ 6 &  0.71  &  9 &  0 &  1 &  g, G10 \\
HVC 26.11+45.88+163 & 257994 & 155354.0+144148 &  \ 146 \ 163 \ 232 \ 217 &  23 ( 3) &  12 $\times$ \ 7 &  2.04  & 22 &  2 &  8 &  g \\
HVC 26.01+45.52+161 & 257956 & 155507.5+142929 &  \ 144 \ 161 \ 230 \ 215 &  25 ( 6) & \  8 $\times$ \ 6 &  1.54  & 14 &  2 &  8 &  g \\
HVC 29.55+43.88+175 & 268067 & 160529.4+160912 &  \ 158 \ 175 \ 255 \ 244 &  37 (11) &  10 $\times$ \ 6 &  1.91  & 20 &  2 &  6 &  g, G10 \\
HVC 28.07+43.42+150 & 268069 & 160532.6+145920 &  \ 132 \ 150 \ 227 \ 214 &  29 ( 4) &   10 $\times$ \ 5 &  1.15 & 11 &  0 & 10 &  g, G10 \\
HVC 28.47+43.13+177 & 268070 & 160707.0+150831 &  \ 160 \ 177 \ 255 \ 243 &  20 ( 3) &  17 $\times$ \ 9 &  1.48  & 11 &  2 &  6 &  g, G10 \\
HVC 28.03+41.54+127 & 268071 & 161236.8+141226 &  \ 109 \ 127 \ 206 \ 194 &  62 (15) &  12 $\times$ \ 7 &  2.67  & 18 &  1 &  8 &  g \\
HVC 28.66+40.38+125 & 268072 & 161745.3+141036 &  \ 108 \ 125 \ 208 \ 197 &  42 ( 5) &  16 $\times$ \ 9 &  3.17  & 21 &  3 &  7 &  g \\
HVC 19.13+35.24-123 & 268213 & 162235.7+050848 &  -139 -123 \ -63 \ -81 &  17 ( 1) &  12 $\times$  10 &  2.83  & 22 &  0 &  7 &  g, G10, S12 \\
HVC 27.86+38.25+124\tablenotemark{$\bigstar$} & 268074 & 162443.4+124412 &  \ 107 \ 124 \ 207 \ 197 &  23 ( 4) &  11 $\times$ \ 9 &  1.28  & 13 &  2 &  4 &  g \\
HVC 84.01-17.95-311 & 310851 & 215406.2+311249 &  -324 -311 \ -98 \ -21 &   21 ( 4) &  26 $\times$ 14 &  2.60  &  17  &   0 &  5 &  g\\
HVC 82.91-20.46-426 & 310865 & 215802.9+283735 &  -439 -426 -217 -140  &  22 ( 1) &  12 $\times$ \ 6 &  0.99  &  10 &   0 & 17 &  g, S12\\
HVC 80.69-23.84-334 & 321318 & 220100.7+244404 &  -345 -334 -131  -55 &   23 ( 1) &  18 $\times$ \ 9 & 1.47  & 13  &  0 &  5 &  g\\
HVC 86.18-21.32-277 & 321455 & 221121.8+295402 &  -288 -277 \ -68 \  \ 10 &  17 ( 1) &  13 $\times$ \ 7 &  1.76  & 15 &  0 &  5 &  g, O \\
HVC 82.91-25.55-291 & 321320 & 221238.6+244311 &  -302 -291 \ -90 \ -13 &  24 ( 2) &  15 $\times$ \ 6 &  1.31  & 13 &  0 &  7 &  g, O \\
HVC 84.61-26.89-330 & 321351 & 222134.4+243638 &  -341 -330 -130 \ -53  &  21   ( 4) &  13 $\times$ 11 &  1.03  &  9 &   0 & 8 &  g, O \\
HVC 92.53-23.02-311 & 321457 & 223823.4+315257 &  -321 -311 -104 \ -23 &  28 ( 2) &  19 $\times$ \ 9 &  1.68  & 12 &  0 &  5 &  g, O \\
HVC 87.35-39.78-454\tablenotemark{a} & 334256 & 230056.4+152014 &  -461 -454 -282 -206 &  26 ( 4) &  11 $\times$ \ 8 &  1.57  & 16 &  0 &  1 &  g, O \\
HVC 88.15-39.37-445\tablenotemark{a} & 334257 & 230211.3+160048 &  -452 -445 -271 -195 &  22 (11) &  12 $\times$ \ 4 &  0.68  &  10 &  0 &  4 &  g, O \\
HVC108.98-31.85-328 & 333613 & 235658.8+293235 &  -333 -328 -147 \ -64 &   19 ( 2)  & 13 $\times$ \ 5 &  0.55 &  8  &  1 & 19 &  g, O \\
HVC109.07-31.59-324 & 333494 & 235702.1+294846 &  -329 -324 -143 \ -60 &   17 ( 5) &  12 $\times$ \ 7 &   1.80 &  23 &  1 &  19 &  g, O\\
\hline
\enddata
\tablenotetext{$\bigstar$} {Part of the extremely isolated MIS subsample}
\tablenotetext{a} {Not included in the \a40 catalog}
\tablenotetext{b} {Also included in the compact cloud catalog of \cite{2012ApJ...758...44S}}
\tablenotetext{c} {Possible kinematic association with larger structure}
\end{deluxetable*}

\begin{figure}[t]
\begin{center}
\includegraphics[trim=1cm 0cm 0cm 0cm,width=0.9\linewidth]{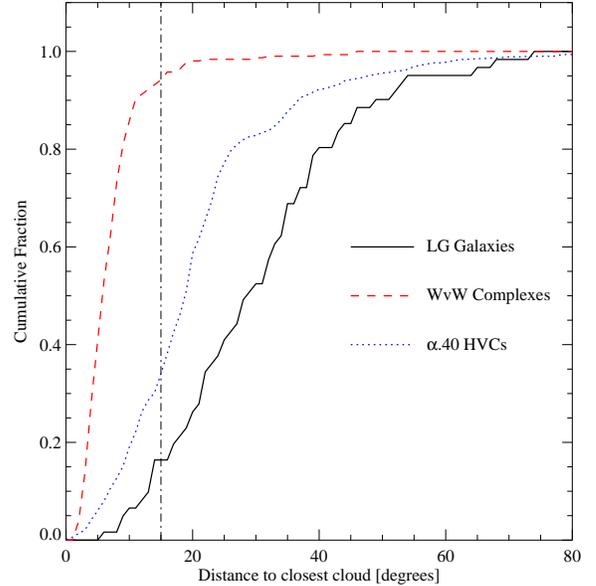}
\caption{The relative isolation of LG galaxies and \a40 HVCs from
the large HVC complexes of WvW.
The x-axis is the distance to the closest WvW cloud in a complex calculated using Equation \ref{eq:dist}.
The y-axis shows the fraction of objects that have their nearest neighbor at that distance or nearer.
The separation of WvW clouds within complexes from each other is shown by the dashed line (red in the online version).
The isolation of the LG galaxies is shown by the solid line (black in the online version), 
and the \a40 HVCs are shown for reference with the dotted line (blue in the online version).
The dot-dash line indicates our chosen isolation criterion of $D=15$\dg.
The majority of clouds in complexes are within 15\dg\ of their nearest neighbor, although there
is a smaller tail extending to 25\dg.
The majority of LG galaxies are located further than 15\dg\ from a cloud in a complex,
making this a good isolation criterion.
This isolation criterion removes $\sim$30\% of the \a40 HVCs from consideration as UCHVCs,
but further isolation criteria are clearly necessary.
}
\label{fig:iso_wvw}
\end{center}
\end{figure}

Our second isolation criterion is that the UCHVCs must be well separated from previously known HVC complexes. 
We compare the UCHVCs to the updated catalog of \citet[B. Wakker, private communication 2012; hereafter WvW]{1991A&A...250..509W}.
The WvW catalog includes 617 clouds, of which 393 are classified as belonging to 20 large complexes;
the other clouds are classified into populations based on their spatial coordinates and velocity. 
For defining isolation, we only consider the WvW clouds which are part of a larger complex.
The distance of a UCHVC from another cloud in degrees can be quantified via:
\begin{equation}\label{eq:dist}
D = \sqrt{ \theta^2 + (f \delta v)^2},
\end{equation}
where $\theta$ is the angular separation in degrees, $\delta v$ is the velocity difference in \kms\ between two clouds, and $f$ 
is a conversion factor that parameterizes the significance we ascribe to the angular separation between two clouds 
versus their difference in velocity in determining whether they are associated with each other.
Following \citet{2012ApJ...758...44S} and \citet{2008ApJ...674..227P}, 
we adopt $f=0.5$\dg/\kms\ as the weighting for the velocity separation for large scale HVC structure.
Figure \ref{fig:iso_wvw} illustrates our determination of the isolation criterion
for deciding if the UCHVCs are separated from the WvW complexes.
 The isolation criterion was determined by 
comparing the separation of clouds within WvW complexes to the separation of 
LG galaxies 
from the nearest WvW cloud in a complex.
The x-axis shows the distance to the nearest WvW cloud in a complex and the y-axis shows 
the fraction of objects whose closest neighbor is at that distance or closer (cumulative fraction).
Ninety percent of WvW clouds in complexes are closer than 15\dg\ to their nearest neighbor in the complex; more than eighty percent of LG galaxies are located further than 15\dg\ from the nearest WvW cloud in a  complex.
Hence we determine to use this value as our cutoff, shown by the dot-dash line in Figure \ref{fig:iso_wvw}.
We note that is a more generous criteria than that of \cite{2012ApJ...758...44S} and \citet{2008ApJ...674..227P} who adopt $D=25$\dg\ as an isolation criterion;
in Section \ref{sec:hvc} we examine this intermediate distance and determine it does not substantially affect our catalog.

\begin{figure}
\begin{center}
\includegraphics[trim=1.5cm 0cm .5cm 0cm,width=0.9\linewidth]{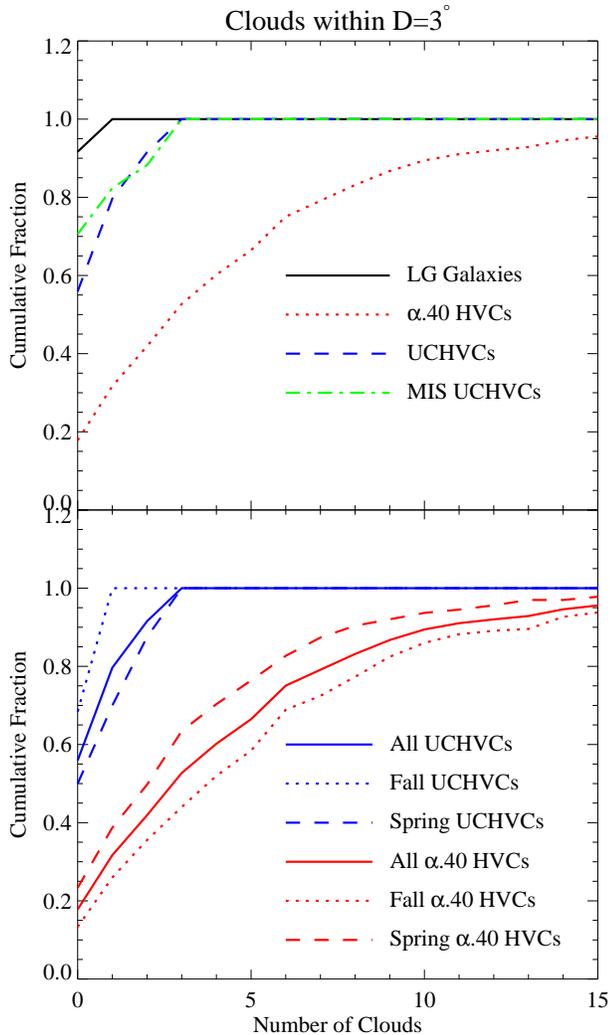}
\caption{
The x-axis is the number of \a40 HVCs within $D=3$\dg,
 where
the distance is calculated from Equation \ref{eq:dist} with $f=0.2$ \dg/\kms,
and the y-axis is the  fraction of UCHVCs with that number of neighbors or fewer.
The top panel shows the relative isolation of LG galaxies (solid line, black in online version),
UCHVCs (dashed line, blue in online version), MIS UCHVCs (dot-dash line, green in online version),
and general \a40 HVCs (dotted line, red in online version).
The LG galaxies have no more than one \a40 HVC within $D=3$\dg;
the criteria for the UCHVCs is slightly relaxed to not more than 3 neighbors.
The \a40 HVCs are shown for reference;
a majority of the \a40 HVCs fail this isolation criteria.
In the bottom panel, we compare the spring (dashed line) and fall populations (dotted line) 
of the UCHVCs (blue in the online version),
with the \a40 HVCs shown for references (red in the online version). 
}
\label{fig:iso_uchvcs}
\end{center}
\end{figure}

In addition, we institute a third isolation criterion based on HVC structure uncovered by ALFALFA. 
This structure is generally much smaller than previously known HVC structure; 
as can be seen in Figure \ref{fig:compare1_9} most \a40 HVCs are less than one degree in size while the sizes of the 
HVCs in the WvW catalog are several to tens of degrees\footnote{As an extragalactic survey, 
ALFALFA was not designed to detect sources with sizes $\gtrsim 1$\dg; 
the commensal GALFA-HI survey which processes the signal independently does that \citep[e.g.][]{2011ApJS..194...20P}.}.
For this reason, we use $f=0.2$\dg/\kms\ in Equation \ref{eq:dist} when calculating isolation from HVC 
structure within the ALFALFA survey. 
The top panel of Figure \ref{fig:iso_uchvcs} shows the final isolation criterion for UCHVCs
 and compares the UCHVCs to
LG galaxies and the general HVC detections within the \a40 survey.
We require that the UCHVCs have no more than three neighbors within $D=3$\dg. 
This is a generous criterion as the LG galaxies have at most one neighbor within this distance.
We wish to include all potential minihalo candidates and inspection indicates that allowing
three neighbors includes all the sources
that would be classified by eye as isolated.
In the bottom panel of Figure \ref{fig:iso_uchvcs} 
we explore the differences between the spring and fall populations of the UCHVCs.
The fall sky appears to show more isolation on this scale with the UCHVCs having either one or no
 neighbors;
in fact, this is a result of the prominent HVC structure in the fall sky.
Clouds in the fall sky are either part of a larger structure or have no (or one) neighbors
within $D=3$\dg.
Comparing to the general \a40 HVC population shows the prevalence of HVC structure in the 
fall sky with the fall HVCs generally having more neighbors than the spring HVCs.

\begin{figure}
\begin{center}
\includegraphics[trim=1.5cm 0cm .5cm 0cm,width=0.9\linewidth]{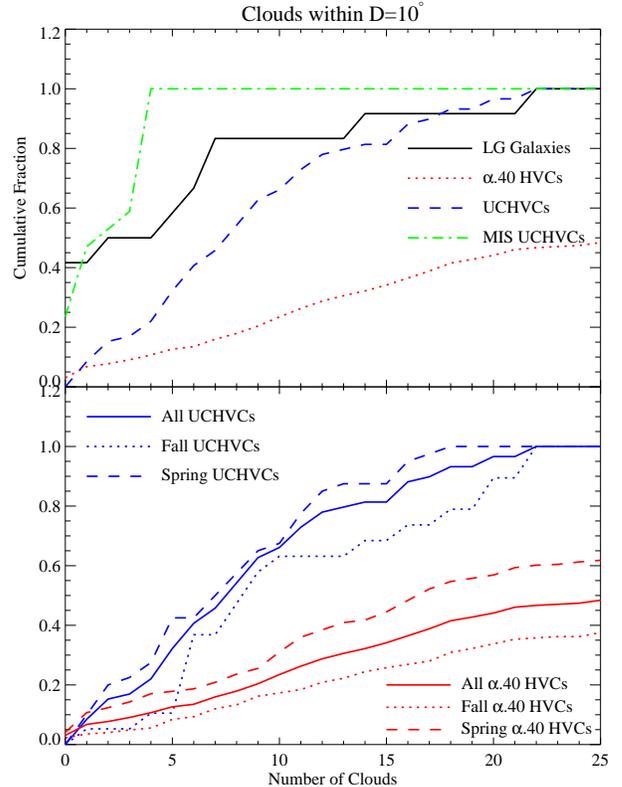}
\caption{
The x-axis is the number of \a40 HVCs within $D=10$\dg,
 where
the distance is calculated from Equation \ref{eq:dist} with $f=0.2$ \dg/\kms,
and the y-axis is the  fraction of UCHVCs with that number of neighbors or fewer.
The top panel shows the relative isolation at this larger distance scale
of LG galaxies (solid line, black in online version),
UCHVCs (dashed line, blue in online version), MIS UCHVCs (dot-dash line, green in online version),
and general \a40 HVCs (dotted line, red in online version).
At this distance scale, the UCHVCs and LG galaxies have similar behavior.
We define a most-isolated subsample (MIS) of UCHVCs which are still isolated
with no more than 3 neighbors on this larger scale.
The MIS UCHVCs are even more isolated than the LG galaxies on this larger scale.
In the bottom panel, we compare the spring (dashed line) and fall populations (dotted line) 
of the UCHVCs (blue in the online version),
with the \a40 HVCs shown for references (red in the online version). 
}
\label{fig:iso_mis}
\end{center}
\end{figure}

We note that with this criterion, only clouds with central velocities
 within 15 \kms\ of the UCHVC can be considered as neighbors. Given that the 
median velocity width of the UCHVCS is 23 \kms, there is a possibility that this
isolation criterion could leave our sources kinematically confused.
Our first isolation criterion accounts for this through the examination of the UCHVCs
for association with other clouds. In order to verify this, we examine the effect
of changing the velocity weighting factor to $f=0.05$\dg/\kms. This expands
the velocity selection to 60 \kms, almost three times the median FWHM of the clouds.
We examine the number of clouds within 3\dg\ of the UCHVCs using
this different value of $f$ and find that the UCHVCs still have very few neighbors
with this modified distance estimate.
In fact, seventy-five percent of the UCHVCs still meet the criterion of three or fewer neighbors
even when the expanded velocity space is considered.
We examined the nine UCHVCs with more than five neighbors and note that three of them
may possibly be kinematically associated with larger structure;
we mark these UCHVCs in Table \ref{tab:uchvc}.

HVC structure often exists on scales much larger than 3\dg; while the UCHVCs are examined for 
obvious connection to larger structure and excluded in that case, 
we still wish to define a more isolated subsample. 
As the best subsample to represent HI sources associated with minihalo candidates, we define a
``most-isolated'' subsample (MIS) of UCHVCs with no more than than 4 neighbors within $D=10$\dg.
The top panel of Figure \ref{fig:iso_mis} 
shows the number of neighboring clouds within $D=10$\dg\ for the UCHVCs, 
the MIS UCHVCs, LG galaxies and \a40 HVCs.
On this large scale, the MIS UCHVCs are generally more isolated than even the LG galaxies.
We do note that the \a40 footprint means that we are not generally probing to a full 10\dg\ 
in all directions around a given cloud; 
increasing coverage of the ALFALFA survey may change the classification of a cloud in the future.
In fact,
two sources in the fall $\delta = +15$\dg\ strip meet the MIS criteria but we exclude them from this subsample as determining isolation out to 10\dg\ for sources in an isolated 2\dg\ wide strip is problematic.
We will revisit these two specific sources and the classification of the MIS UCHVCs 
in general with increased ALFALFA coverage in future work.
In the bottom panel of Figure \ref{fig:iso_mis}, we again examine the difference 
between the fall and spring population.
Here, the prominent HVC structure in the fall sky is apparent with 
many of the fall UCHVCs having a large number of
neighbors out to a distance of 10\dg.
There is also a strong difference evident between the UCHVC and \a40 HVC population
with over half of the \a40 HVCs having more than 20 neighbors at $D=10$\dg;
this indicates the utility of our first isolation criterion of inspecting sources
for connection to large scale structure.

\section{Catalog}\label{sec:catalog}

\subsection{Presentation of Catalog}
In Table \ref{tab:uchvc} we present the UCHVCs; there are \ntot\ sources total: \nspring\ in the spring \a40 sky and \nfall\ in the fall sky. Of the \ntot\ UCHVCs, \nmh\ are identified as being in the 
most-isolated subsample, all of which are in the spring sky.
The spring sky samples the outer regions of the LG where the expected density 
for dark matter halos may be lower
but the environment is safer for gas-bearing minihalos than near the MW or M31.
The fall sky samples the LG near M31 and includes the presence of a large amount of HVC
structure, including the Magellanic Stream (see Section \ref{sec:hvc} for a further discussion).
We indicate those UCHVCs that are part of the original sample of UCHVCs discussed by G10 with a G10 in the notes column and those 
UCHVCs that lie outside the area considered by G10 with an `O'.
Figure \ref{fig:contour} shows maps of all the UCHVCs with contours in units
of column density of HI ($N_{HI}$ in atoms cm$^{-2}$), representing the sum total
of HI content along the line of sight; these plots represent the data from which all
the parameters listed in Table \ref{tab:uchvc} are derived.
The minimum contour level is given in the figure and subsequent contour levels increase by factors of $\sqrt{2}$.
We plot the contours in values of $N_{HI}$ to demonstrate that the peak column density value is higher
than the average value calculated later (see Section \ref{sec:inf}).
However, we emphasize, that since these clouds are barely resolved by the Arecibo beam, the column density contour 
values are only approximate and the average values are more robust;
to accurately map the distribution of HI will require synthesis observations that provide a smaller beam.
Column density values can be derived from the brightness temperature via:
\begin{equation}
N_{HI} = 1.823 \times 10^{18} \int T_B\, dv \, \,  [{\rm cm}^{-2}].
\end{equation}
In simple cases, the brightness temperature is related to the flux density at 21cm via:
\begin{equation}
 T_B  = \frac{606}{\theta^2} S
\end{equation}
where $\theta$ is the (circular) beam in arcseconds and $S$ the flux in mJy/beam.

\begin{figure*}
\begin{center}
\includegraphics[keepaspectratio,width=0.8\linewidth]{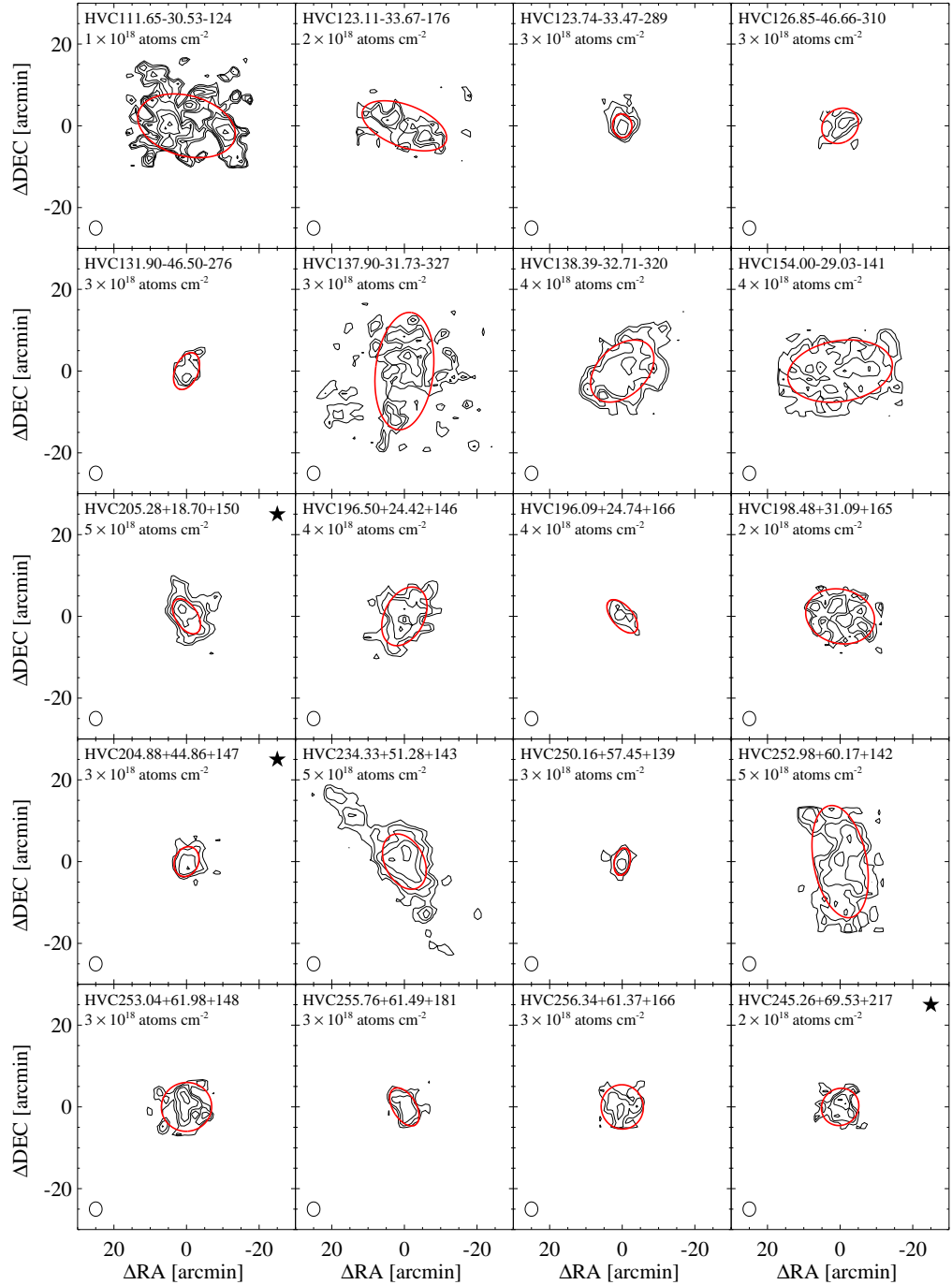}
\caption{
Maps of the HI column density 
of the UCHVCs derived from ALFALFA spectral grids.
 Starred figures indicate membership in the most-isolated subsample.
Ellipses (red in the online version) represent the measured half-power level.
The 3\arcmin.5 circular beam is shown in the lower left corner of all plots.
The lowest contour level is listed in the upper left corner of each plot;
subsequent contours increase by factors of $\sqrt{2}$.
}
\label{fig:contour}
\end{center}
\end{figure*}

\addtocounter{figure}{-1}
\begin{figure*}
\begin{center}
\includegraphics[keepaspectratio,width=0.8\linewidth]{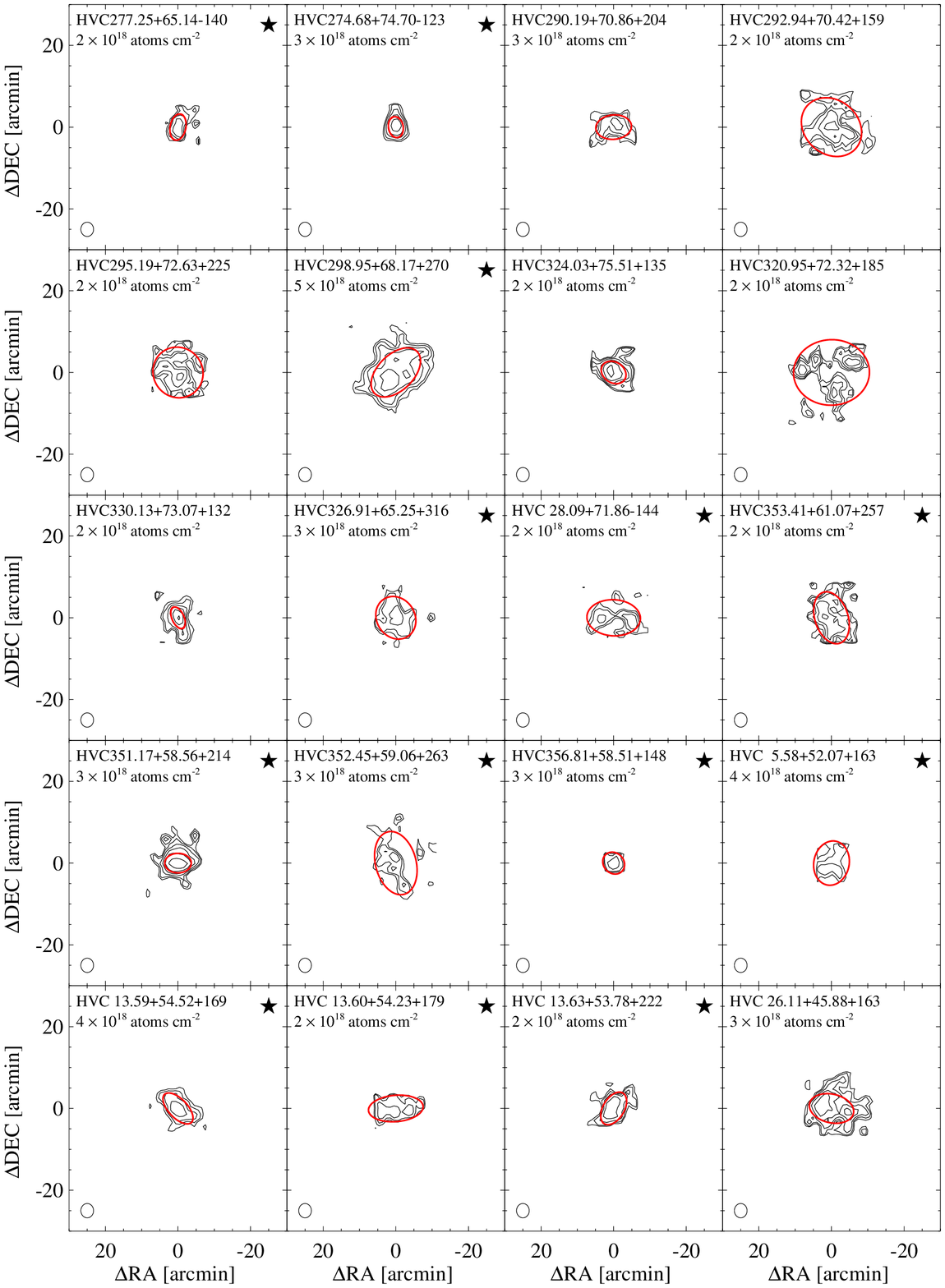}
\caption{Continued}
\label{fig:contour2}
\end{center}
\end{figure*}

\addtocounter{figure}{-1}
\begin{figure*}
\begin{center}
\includegraphics[keepaspectratio,width=0.8\linewidth]{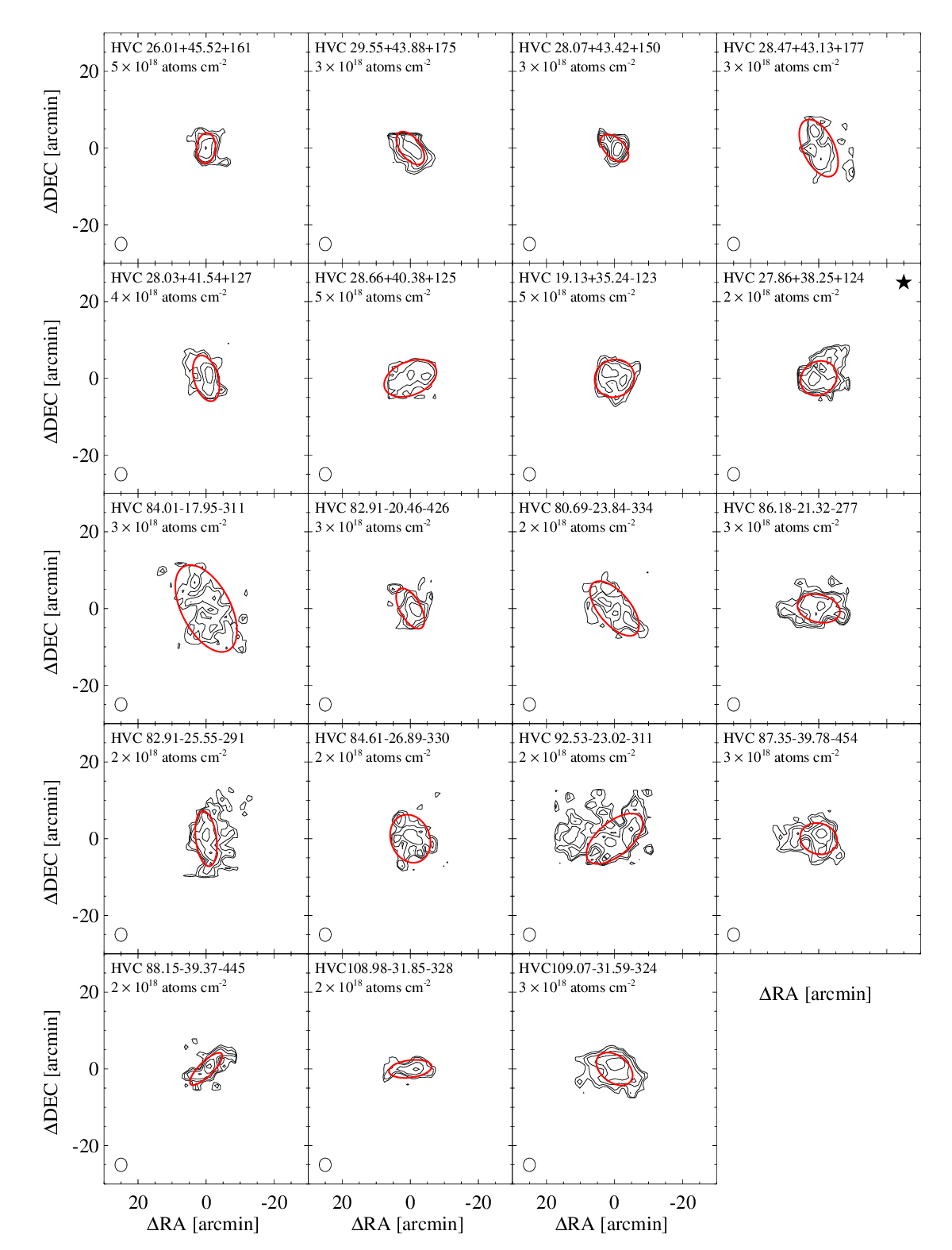}
\caption{Continued}
\label{fig:contour3}
\end{center}
\end{figure*}

The columns of the tables are as follows:
\begin{itemize}
\item Col. 1:      Source name, in the traditional form for HVCs, obtained from the galactic
          coordinates at the nominal cloud center and the \vlsr\ of the cloud, e.g. HVC111.65-30.53-124 has
          $l=$111.65\dg, $b=$-30.53\dg, and \vlsr = -124 \kms.
\item Col. 2:     Identification number in the Arecibo General Catalog (AGC), an internal database maintained
by MH and RG,
           included to ease cross--reference with
          our archival system and the \a40 catalog.
\item Col. 3:     Equatorial coordinates of the centroid, epoch J2000. Typical errors are less than 1\arcmin.
\item Col. 4:     Sequentially, we list heliocentric velocity, velocity in the local standard of rest frame (LSR; assumed solar
          motion of 20 \kms ~towards $l=57^\circ$, $b=25^\circ$), 
velocity in the Galactic standard of rest frame (GSR;
          $V_{gsr}=V_{lsr}+225 \sin l\cos b$, with both velocities in \kms), and the
          velocity with respect to the LG reference frame from \cite{1996AJ....111..794K}.
\item Col. 5:     HI line full width at half maximum ($W_{50}$), with estimated measurement error in brackets.
          The notes column indicates the method of measurement: a gaussian fit or linear single peaks fit to the sides of the profile.
\item Col. 6:     Estimate of the cloud  major and minor diameters, in arcminutes. Sizes are
          measured at approximately the level encircling half the total flux density. In many
          cases, the outer contours are more elongated than indicated by the ratio $a\times b$.
                The half-power ellipses are also shown in the HI column density contour plots in Figure \ref{fig:contour}.
\item Col. 7:     Flux density integral ($S_{21}$), in Jy \kms.
\item Col. 8:     Signal--to-noise ratio (S/N) of the line, defined as 
\begin{equation}
S/N = (\frac{1000 S_{21}}{W_{50}}) \frac{w_{smo}^{1/2}}{\sigma_{rms}},
\end{equation}
where $S_{21}$ is the integrated flux density in Jy \kms, as listed in Column 7; the ratio 1000$S_{21}$/$W_{50}$ is the mean flux density across the feature in mJy; 
$w_{smo}$ is  $W_{50}$/(2 $\times$ 10), a smoothing width, and $\sigma_{rms}$ is the rms noise figure across the spectrum measured in mJy.
More details on the S/N calculation are available in \cite{2011AJ....142..170H}.
\item Col. 9: The number of \a40 HVC neighbors within $D=3$\dg\ (for $f=0.2$\dg/\kms)
\item Col. 10: The number of \a40 HVC neighbors within $D=10$\dg\ (for $f=0.2$\dg/\kms)
\item Col. 11:       Notes column. For each source there is either a `g' or `p' indicating the method used 
(gaussian or single peaks fit) to measure $W_{50}$. 
Sources considered by G10 are indicated with a `G10' in the notes column. 
Sources that are outside the footprint considered in G10 are marked with a `O'. 
The UCHVCs that are also in the GALFA compact cloud catalog of \citet{2012ApJ...758...44S} are indicated with a `S12'.

\end{itemize}


\subsection{Comparison to G10}
For completeness, we include in Table \ref{tab:fail} the UCHVCs 
that were considered by G10 but 
do not meet the stricter selection criteria used here.
The clouds from G10 can fail any of the criteria: S/N, isolation or \vlsr\ limits. 
The notes column indicates the reason a G10 cloud is not included here.
The sources with S/N \textless\ 8 will be considered in future work when we extend the UCHVC catalog to lower S/N values after assessing reliability and completeness. 
In addition, we will extend the catalog to velocities including the Galactic hydrogen.
It should be noted that the three sources that do not meet the isolation criteria only barely fail.
Two sources have one and two more neighbors than allowed, respectively, and the third sources is
 excluded based on examination of large scale structure.
These sources could still be good minihalo candidates.

\begin{deluxetable*}
{lcccccccccl} 
\tablewidth{0pt}
\tabletypesize{\scriptsize}
\tablecaption{UCHVCs from G10 that Fail UCHVC Criteria  \label{tab:fail}}
\tablehead{
\colhead{Source}  & AGC    & \colhead{R.A.+ Dec.} &
\colhead{cz$_\odot$ $V_{lsr}$ $V_{gsr}$ $V_{LG}$} & \colhead{$W_{50}(\epsilon_w)$}  &
\colhead{$a\times b$} & \colhead{$S_{21}$}      & \colhead{S/N} &  \colhead{$N_3$} & \colhead{$N_{10}$} & Reason  \\
{} & {} & J2000 &  \kms & \kms & \arcmin & Jy \kms & & & &
}
\startdata
HVC244.51+53.41+160 & 208424 & 104850.1+050419 &  \ 164 \ 160 \ \ 39 \ -18 &   19 ( 3) & 16 $\times$ 12 &   1.03 &   7 &  0 &  9 & S/N \\
HVC249.03+57.58+178 & 219213 & 110813.6+055725 &  \ 179 \ 176  \ 64 \ \ \ \ 5 &   19 ( 2) & 12 $\times$ \ 9 &   0.67 &   7&  0 &  8 & S/N \\
HVC247.19+70.29+247 & 215418 & 114418.2+150509 &  \ 246 \ 247 \ 177 \ 129 &   30 (10) &  10 $\times$ \ 8 &   0.54 &   7 & 0 &  1& S/N  \\
HVC290.37+66.23-115 & 227983 & 123116.7+035044 &   -118 -114 -199 -259 & 20 (5) & \ 6$\times$ \ 4& 0.44 & 9 & 0 &  3& Velocity\\
HVC298.30+72.91+185 & 226171 & 124557.2+100518 & \ 180 \ 185 \ 127 \ \ 75 & 25 (3) & \ 5 $\times$ \ 4 & 0.57 & 9 &   5 & 21& Isolation\\
HVC299.62+67.65+326 & 227988 & 124619.1+044923 &  \ 323 \ 327 \ 253 \ 195 &   39 (13) & 14 $\times$ \ 7 &   0.76 &   6 & 0 &  0&  S/N\\
HVC314.57+74.80+218 & 238626 & 130351.1+121223 &  \ 211 \ 218 \ 176 \ 127 &   36 (13) & \ 5 $\times$ \ 3 &   0.35 &   5 & 0 & 17&  S/N\\
HVC  8.88+62.16+281 & 249538 & 143531.7+133126 &  \ 269 \ 282 \ 298 \ 264 &   18 ( 6) & \ 4 $\times$ \ 3 &   0.22 &   4 &     0  & 4 & S/N\\
HVC  7.64+57.83-128 & 249248 & 144844.6+103510 &  -142 -128 -112 -147 &   22 (1) & 25$\times$ \ 5  &   1.83 &  16  &  0 & 42& Isolation\\
HVC 15.11+45.54-148 & 258474 & 154035.2+074334 & -163 -147 -106 -132 & 27 (1) & \ 7 $\times$ \ 5 & 0.68 & 9 &  4 & 19& Isolation\\
\hline
\enddata
\end{deluxetable*}

\subsection{Properties of the UCHVCs}

\begin{figure}
\begin{center}
\includegraphics[keepaspectratio,width=0.9\linewidth]{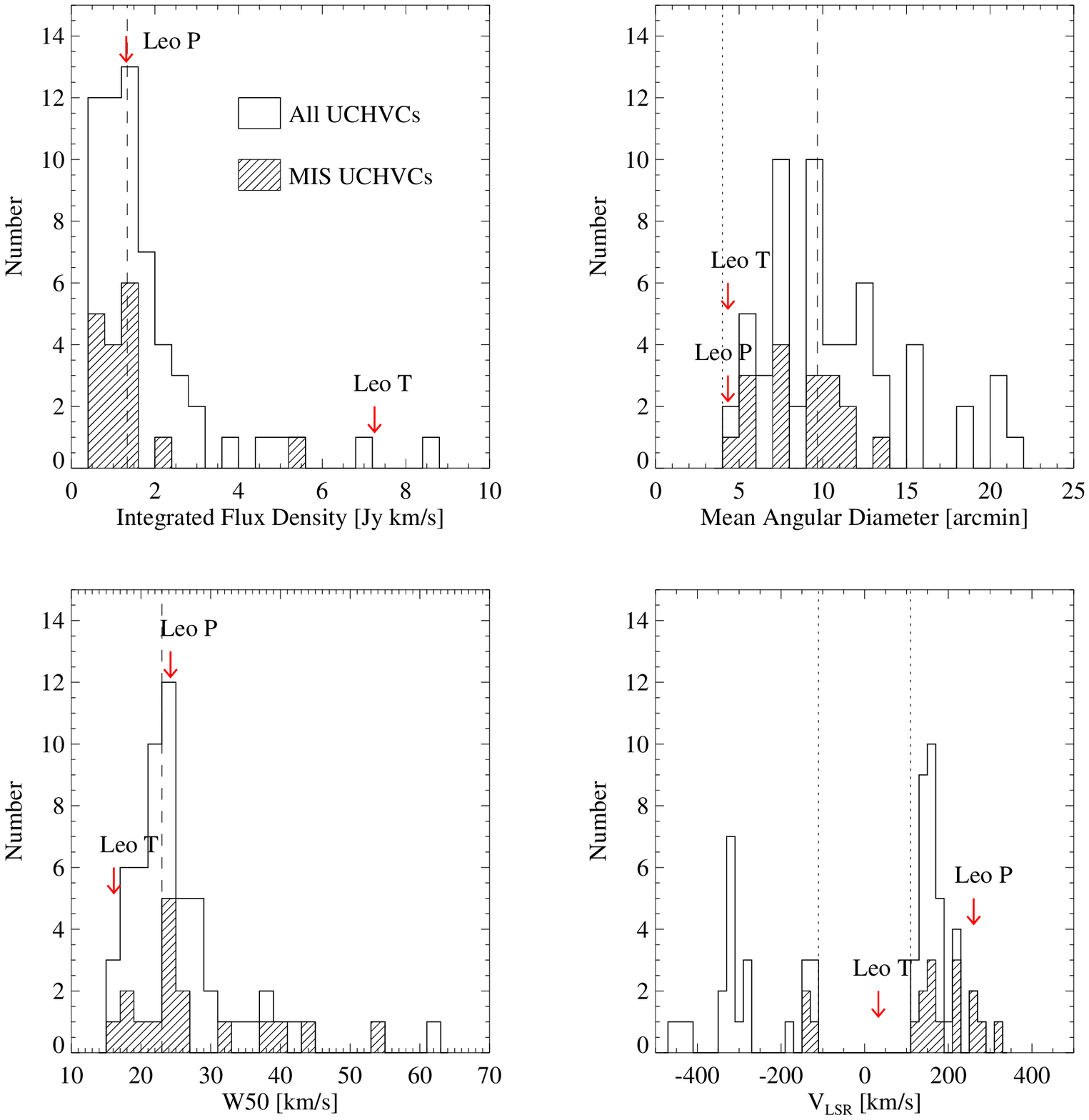}
\caption{Histograms of measured properties for the UCHVCs. Hashed histograms indicate the most-isolated subsample.
The measured values for Leo T and Leo P from the ALFALFA data are indicated with arrows (red in the online version).
The dashed lines are the median values of the UCHVCs;
the most-isolated subsample has a slightly lower median flux density value and
identical median values for the HI size and $W_{50}$.
The dotted lines indicate observational boundaries.
In the upper right panel, the dotted line indicates the smallest structure that can be resolved by Arecibo,
and in the the bottom right panel the dotted lines indicate the velocity selection criterion.
}
\label{fig:meas_hist}
\end{center}
\end{figure}

Figure \ref{fig:meas_hist} shows the distribution of measured properties for the \a40 UCHVCs and the most-isolated subsample: 
 integrated flux density ($S_{21}$), average angular diameter ($\bar a = \sqrt{a b}$), velocity FWHM ($W_{50}$), and \vlsr. 
The UCHVCs have integrated flux densities of $\sim$0.66-8.55 Jy \kms, 
with the vast majority having integrated flux densities below 3.5 Jy \kms\ 
and a median flux density of \fmedian\ Jy \kms.
The singly hatched histograms are the UCHVCs in the most-isolated subsample.
Note that the range of values for the MIS UCHVCs is similar to the larger UCHVC population, and the median values are essentially identical.
The UCHVCs range in average diameter from essentially unresolved ($\sim$4\arcmin) to just over 20\arcmin\ in size, 
with the vast majority less than 16\arcmin\ in size and a median size of \hsizemedian \arcmin.
We note that there does appear to be a break in population based on size with UCHVCs clustered with HI diameters \textless\ 16\arcmin\ in size and a tail of a population extending to larger sizes (including objects with HI diameters \textgreater\ 30\arcmin\ not included in this work). 
We will explore this break in HI size in the HVC population in future work with a larger survey area.
The $W_{50}$ values are centered around 15-30 \kms\ with a few UCHVCs having widths extending up to 70 \kms; the median linewidth is \wmedian\ \kms.
There are clouds whose velocities cluster near both \vlsr\ $\pm$120 \kms,
with a much stronger clustering of positive velocity clouds. 
However, when the MIS UCHVCs are considered, this clustering disappears.
The vast majority of negative velocity clouds are also excluded from the MIS UCHVCs; the negative velocity clouds
are predominantly in the fall sky, where large scale HI structure is much more prevalent, preventing the inclusion
of any UCHVCs into the most-isolated subsample.

\subsection{Inferred Cloud Parameters}\label{sec:inf}

Given the observed properties of the UCHVCs, integrated flux density ($S_{21}$, Jy \kms), average angular diameter ($\bar a = \sqrt{a b}$, arcminutes) and
velocity width ($W_{50}$, \kms), it is straightforward to derive some simple properties of the UCHVCs, modulo the unknown distance $d$ (in Mpc), with the assumption that the clouds are optically thin. Sequentially, below we derive the mean
atomic density, mean column density, HI mass, indicative dynamical mass within the HI extent, and HI diameter.
\begin{eqnarray}
\bar n_{HI} [\mathrm{atoms\ cm}^{-2}] &  =  &  0.74 ~S_{21}  ~\bar a^{-3} ~d^{-1} \,\,\,\,{\rm cm}^{-3} \label{eq:dens}\\
\bar N_{HI} [\mathrm{atoms\ cm}^{-2}]&  = & 4.4\times 10^{20} ~\bar a^{-2} ~S_{21} \,\,\,\,{\rm cm}^{-2} \label{eq:coldens}\\
M_{HI} [M_\odot] & = & 2.356\times 10^5 ~S_{21} ~d^2   \label{eq:HImass}\\
M_{dyn} [M_\odot] & = & 6.2\times 10^3 ~\bar a~W_{50}^2 ~ d  \label{eq:mctot}\\
D_{HI} [\mathrm{kpc}] & = & 0.29 ~\bar a~d  \label{eq:radius}
\end{eqnarray}
Of these derived properties, $\bar N_{HI}$ is especially noteworthy as it does not depend on the distance.
It should be noted that the column density values derived here are average values based on the global properties of the UCHVCs,
in contrast to the approximation of spatially-resolved column density contours in Figure \ref{fig:contour}.
Due to the large beam size of Arecibo, these values represent underestimates of the peak values of the clouds.
We note that the dynamical mass is an indicative mass dynamical mass only. In addition to the uncertainty in the distance of the UCHVCs, the contribution to the linewidths of the UCHVCs from thermal broadening is unknown. For a range of reasonable temperatures, the thermal broadening can range from 16-21 \kms. For the clouds with the largest linewidths, the thermal broadening contribution (when accounted for in quadrature) may be negligible, while the narrowest clouds may be fully thermally supported. However, they could still have large-scale motions on the order of the thermal broadening, or less. 
For example, Leo P has a linewidth of 24 \kms\ and a rotational velocity of 9 \kms, uncorrected for disk inclination \citep{LeoP_HI}.
To derive accurate dynamical masses will require higher resolution HI images in which evidence of large scale motions can be discerned (and, of course, distance information).


In Table \ref{tab:parms}, we summarize the inferred properties of the UCHVCs.
The columns of the table are as follows:
\begin{itemize}
\item Col. 1 and 2: source id as in Table \ref{tab:uchvc}
\item Col 3: HI diameter in kpc at $d=1$ Mpc (Eqn. \ref{eq:radius})
\item Col 4: log of the mean atomic HI density at $d=1$ Mpc, in cm$^{-3}$ (Eqn. \ref{eq:dens})
\item Col 5: log of the mean HI column density, in cm$^{-2}$ (Eqn. \ref{eq:coldens})
\item Col 6: log of the HI mass at $d=1$ Mpc, in solar units (Eqn. \ref{eq:HImass})
\item Col 7: log of the indicative dynamical mass within $D_{HI}$ at $d=1$ Mpc,
in solar units (Eqn. \ref{eq:mctot})
\end{itemize}

\begin{deluxetable*}{lcccccc} 
\tablewidth{0pt}
\tabletypesize{\scriptsize}
\tablecaption{Inferred Cloud Properties \label{tab:parms}}
\tablehead{
\colhead{Source}& \colhead{AGC}  & \colhead{$D_{HI}$} & 
\colhead{$n_{HI}$} & 
\colhead{$\log~\bar N_{HI}$} & \colhead{$\log M_{HI}$} & 
\colhead{$\log M_{dyn}$}  
 \\
{} & {} & kpc $d$  & cm$^{-3} d^{-1}$ & cm$^{-2}$ & $M_\odot d^{2}$ & $M_\odot d$ 
}
\startdata
HVC111.65-30.53-124 & 103417 & 5.8 & -3.68 & 18.40 & 5.74 & 7.74 \\ 
HVC123.11-33.67-176 & 102992 & 4.6 & -3.62 & 18.35 & 5.48 & 7.64 \\ 
HVC123.74-33.47-289 & 102994 & 1.6 & -2.54 & 18.98 & 5.20 & 7.18 \\ 
HVC126.85-46.66-310 & 749141 & 2.7 & -3.12 & 18.62 & 5.28 & 7.48 \\ 
HVC131.90-46.50-276 & 114574 & 2.2 & -2.95 & 18.72 & 5.22 & 7.54 \\ 
HVC137.90-31.73-327 & 114116 & 6.2 & -3.53 & 18.58 & 5.97 & 8.19 \\ 
HVC138.39-32.71-320 & 114117 & 4.5 & -3.07 & 18.90 & 6.02 & 7.67 \\ 
HVC154.00-29.03-141 & 122836 & 6.0 & -3.25 & 18.85 & 6.21 & 7.97 \\ 
HVC205.28+18.70+150\tablenotemark{$\bigstar$} & 174540 & 2.2 & -2.46 & 19.19 & 5.69 & 7.40 \\ 
HVC196.50+24.42+146 & 174763 & 3.8 & -3.02 & 18.87 & 5.82 & 7.51 \\ 
HVC196.09+24.74+166 & 174764 & 2.2 & -2.94 & 18.71 & 5.19 & 7.43 \\ 
HVC198.48+31.09+165 & 189054 & 4.6 & -3.49 & 18.49 & 5.62 & 7.82 \\ 
HVC204.88+44.86+147\tablenotemark{$\bigstar$} & 198511 & 2.0 & -2.81 & 18.81 & 5.24 & 6.99 \\ 
HVC234.33+51.28+143 & 208315 & 3.6 & -2.72 & 19.15 & 6.07 & 7.49 \\ 
HVC250.16+57.45+139 & 219214 & 1.6 & -2.58 & 18.93 & 5.12 & 7.13 \\ 
HVC252.98+60.17+142 & 219274 & 5.8 & -3.11 & 18.97 & 6.30 & 7.96 \\ 
HVC253.04+61.98+148 & 219276 & 3.7 & -3.14 & 18.74 & 5.69 & 8.01 \\ 
HVC255.76+61.49+181 & 219278 & 2.4 & -2.91 & 18.78 & 5.33 & 7.21 \\ 
HVC256.34+61.37+166 & 219279 & 3.2 & -3.10 & 18.72 & 5.55 & 7.60 \\ 
HVC245.26+69.53+217\tablenotemark{$\bigstar$} & 215417 & 2.8 & -3.24 & 18.52 & 5.22 & 7.24 \\ 
HVC277.25+65.14-140\tablenotemark{$\bigstar$} & 227977 & 1.5 & -2.64 & 18.86 & 5.03 & 7.24 \\ 
HVC274.68+74.70-123\tablenotemark{$\bigstar$} & 226067 & 1.3 & -2.14 & 19.29 & 5.34 & 7.91 \\ 
HVC290.19+70.86+204 & 226165 & 2.2 & -2.83 & 18.83 & 5.33 & 7.32 \\ 
HVC292.94+70.42+159 & 229344 & 4.4 & -3.46 & 18.50 & 5.59 & 7.33 \\ 
HVC295.19+72.63+225 & 226170 & 3.8 & -3.41 & 18.48 & 5.44 & 7.80 \\ 
HVC298.95+68.17+270\tablenotemark{$\bigstar$} & 227987 & 3.5 & -2.62 & 19.23 & 6.12 & 7.70 \\ 
HVC324.03+75.51+135 & 233763 & 1.8 & -2.51 & 19.05 & 5.35 & 7.50 \\ 
HVC320.95+72.32+185 & 233830 & 5.3 & -3.69 & 18.35 & 5.60 & 7.78 \\ 
HVC330.13+73.07+132 & 233831 & 1.2 & -2.23 & 19.17 & 5.17 & 6.83 \\ 
HVC326.91+65.25+316\tablenotemark{$\bigstar$} & 238713 & 3.1 & -3.12 & 18.68 & 5.47 & 7.65 \\ 
HVC 28.09+71.86-144\tablenotemark{$\bigstar$} & 249393 & 3.3 & -3.25 & 18.58 & 5.42 & 8.11 \\ 
HVC353.41+61.07+257\tablenotemark{$\bigstar$} & 249323 & 3.2 & -3.12 & 18.69 & 5.50 & 7.43 \\ 
HVC351.17+58.56+214\tablenotemark{$\bigstar$} & 249282 & 1.7 & -2.29 & 19.26 & 5.53 & 7.77 \\ 
HVC352.45+59.06+263\tablenotemark{$\bigstar$} & 249283 & 3.9 & -3.46 & 18.44 & 5.42 & 7.92 \\ 
HVC356.81+58.51+148\tablenotemark{$\bigstar$} & 249326 & 1.6 & -2.54 & 18.99 & 5.22 & 7.70 \\ 
HVC  5.58+52.07+163\tablenotemark{$\bigstar$} & 258459 & 3.0 & -3.05 & 18.74 & 5.50 & 7.57 \\ 
HVC 13.59+54.52+169\tablenotemark{$\bigstar$} & 258237 & 2.0 & -2.55 & 19.08 & 5.50 & 7.36 \\ 
HVC 13.60+54.23+179\tablenotemark{$\bigstar$} & 258241 & 2.9 & -3.12 & 18.65 & 5.37 & 7.25 \\ 
HVC 13.63+53.78+222\tablenotemark{$\bigstar$} & 258242 & 2.1 & -2.84 & 18.79 & 5.22 & 7.29 \\ 
HVC 26.11+45.88+163 & 257994 & 2.7 & -2.72 & 19.02 & 5.68 & 7.48 \\ 
HVC 26.01+45.52+161 & 257956 & 1.9 & -2.40 & 19.19 & 5.56 & 7.41 \\ 
HVC 29.55+43.88+175 & 268067 & 2.2 & -2.51 & 19.15 & 5.65 & 7.81 \\ 
HVC 28.07+43.42+150 & 268069 & 2.1 & -2.62 & 19.00 & 5.43 & 7.57 \\ 
HVC 28.47+43.13+177 & 268070 & 3.5 & -3.22 & 18.64 & 5.54 & 7.48 \\ 
HVC 28.03+41.54+127 & 268071 & 2.7 & -2.62 & 19.13 & 5.80 & 8.35 \\ 
HVC 28.66+40.38+125 & 268072 & 3.4 & -2.85 & 19.00 & 5.87 & 8.11 \\ 
HVC 19.13+35.24-123 & 268213 & 3.1 & -2.77 & 19.04 & 5.82 & 7.28 \\ 
HVC 27.86+38.25+124\tablenotemark{$\bigstar$} & 268074 & 2.8 & -3.00 & 18.77 & 5.48 & 7.51 \\ 
HVC 84.01-17.95-311 & 310851 & 5.4 & -3.54 & 18.51 & 5.79 & 7.71 \\ 
HVC 82.91-20.46-426 & 310865 & 2.4 & -2.91 & 18.79 & 5.37 & 7.40 \\ 
HVC 80.69-23.84-334 & 321318 & 3.7 & -3.27 & 18.61 & 5.54 & 7.62 \\ 
HVC 86.18-21.32-277 & 321455 & 2.8 & -2.82 & 18.93 & 5.62 & 7.23 \\ 
HVC 82.91-25.55-291 & 321320 & 2.7 & -2.93 & 18.81 & 5.49 & 7.52 \\ 
HVC 84.61-26.89-330 & 321351 & 3.5 & -3.37 & 18.49 & 5.39 & 7.52 \\ 
HVC 92.53-23.02-311 & 321457 & 3.8 & -3.27 & 18.63 & 5.60 & 7.81 \\ 
HVC 87.35-39.78-454 & 334256 & 2.7 & -2.85 & 18.89 & 5.57 & 7.59 \\ 
HVC 88.15-39.37-445 & 334257 & 2.0 & -2.80 & 18.81 & 5.20 & 7.31 \\ 
HVC108.98-31.85-328 & 333613 & 2.2 & -3.02 & 18.63 & 5.11 & 7.23 \\ 
HVC109.07-31.59-324 & 333494 & 2.7 & -2.77 & 18.97 & 5.63 & 7.22 \\ 
\hline
\enddata
\tablenotetext{$\bigstar$} {Part of the extremely isolated MIS subsample}
\end{deluxetable*}

\begin{figure}
\begin{center}
\includegraphics[keepaspectratio,width=0.9\linewidth]{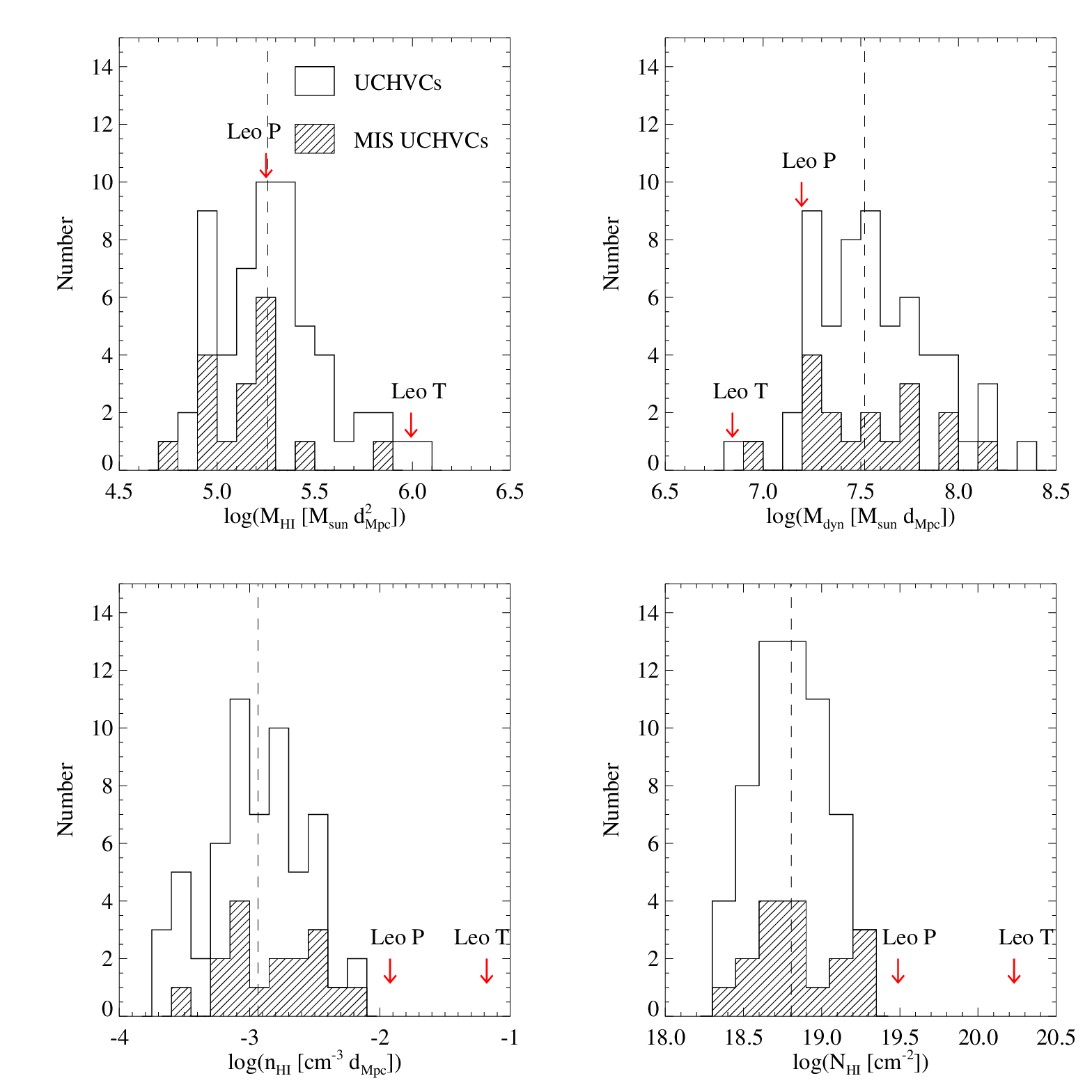}
\caption{The distribution of inferred properties for the UCHVCs. 
Shading and symbols are the same as in Figure \ref{fig:meas_hist}.
The most-isolated subsample has a slightly lower median mass than the full UCHVC sample;
for other properties the median values are equivalent between the two samples. Leo T and Leo P
are shown for comparison.
}
\label{fig:inf_hist}
\end{center}
\end{figure}

The HI masses, dynamical masses, mean atomic densities and mean column densities of
the UCHVCs and the MIS UCHVCs are shown in Figure \ref{fig:inf_hist}.
At a distance of 1 Mpc, the HI masses are around $\sim 10^5-10^6$ \msun\ 
and the dynamical masses are $\sim 10^7-10^8$ \msun.
This would require the UCHVCs to have an ionized envelope of hydrogen or a substantial amount of dark matter in order to be self-gravitating.
As discussed in Section \ref{sec:mhc}, these median properties are a good match to the minihalo models of \cite{2002ApJS..143..419S}.
The median dynamical mass is $10^{7.5}\, d_{Mpc}$ \msun; this is close to the common mass scale of $\sim10^7$\msun\ for 
the UFDs of \cite{2008Natur.454.1096S}.

\section{The UCHVCs as a Distinct Population}\label{sec:other}

While the minihalo hypothesis is intriguing for the UCHVCs, we must carefully consider other possible explanations. 
In this section we examine the possibility of associating the UCHVCs with other cloud populations,
including large HVC complexes, the Magellanic Stream, Galactic halo clouds, and the small cloud populations of the GALFA-HI
survey.

\subsection{The UCHVCs in the Context of Large HVC Complexes}\label{sec:hvc}

The HVC sky contains many large extended structures composed of multiple clouds. 
We explicitly require the UCHVCs to be isolated from the known large scale HVC structure of the WvW catalog.
However, our isolation criterion for separation from WvW complexes is slightly relaxed
in order to avoid excluding potential minihalo candidates.
As can be seen in Figure \ref{fig:iso_wvw}, 
the distance to the nearest cloud within a WvW complex can extend to $D=25$\dg.
As we set our isolation criterion for UCHVCs to a separation of 15\dg\ from WvW clouds in complexes,
we wish
here to  consider the possible association of the UCHVCs with WvW complexes.
In Table \ref{tab:hvc} we list the UCHVCs that are less than 25\dg\ from a WvW complex.
We note that only two UCHVCs in the fall sky (HVC86.18-21.32-277 and HVC87.35-39.78-454) are
more than 25\dg\ from a complex in the WvW catalog;
the other fall HVCs not listed in Table \ref{tab:hvc} are separated by less than 25\dg\ from clouds associated with the Magellanic Stream in the WvW catalog.
Of the \nspring\ spring UCHVCs, seven are potentially associated with known large complexes, the majority of those being with the
WA complex.
While a few of the UCHVCs may be associated with known large complexes, the vast majority are not, as defined by our isolation criterion.

\begin{deluxetable}
{lll} 
\tablewidth{0pt}
\tabletypesize{\scriptsize}
\tablecaption{UCHVCs within $D=25$\dg\ of a WvW complex \label{tab:hvc}}
\tablehead{
\colhead{Complex}  & \colhead{UCHVC}    & \colhead{Distance to closest cloud} \\
{} & {} & {degrees} 
}
\startdata
Complex G  &  HVC111.65-30.53-124 & 20.1 \\
Complex H &  HVC123.11-33.67-176 & 17.9 \\
Complex ACVHV & HVC137.90-31.73-327 & 23.8 \\
              & HVC138.39-32.71-320 & 20.9 \\
Complex ACHV &  HVC154.00-29.03-141 & 15.1\\
Complex WC & HVC205.28+18.70+150 & 24.6\\
Complex WA & HVC234.33+51.28+143 & 16.3 \\
       & HVC250.16+57.45+139 & 19.3 \\
       & HVC252.98+60.17+142 & 21.9 \\
       & HVC253.04+61.98+148 & 24.6 \\
       & HVC256.34+61.37+166 & 24.7 \\
Complex C & HVC 19.13+35.24-123 & 19.4\\
\hline
\enddata
\end{deluxetable}

\subsubsection{Magellanic Stream}\label{sec:ms}
The Magellanic Stream (MS) is an extended HI structure first noted by \cite{1965AJ.....70..552D} and first associated with the Magellanic Clouds by \cite{1974ApJ...190..291M}. 
The MS is generally associated with the disruption of the Magellanic Clouds as they interact with the Milky Way, 
although the exact mechanisms responsible for the MS are an open area of research.
The two main parts of the MS are the Leading Arm (LA), which consists of gas ahead of the Large Magellanic Cloud (LMC) and Small Magellanic Cloud (SMC) in their presumed orbits, and the tail, which consists of the trailing material.
Recently, \citet[hereafter N10]{2010ApJ...723.1618N} presented an extension of the Magellanic Stream (MS), bringing it to over a 200\dg\ length in total.
Given the extent of the MS, 
possible association with the MS must be considered when attempting to understand HVCs of any sort.

\begin{figure}
\begin{center}
\includegraphics[keepaspectratio,trim=2.5cm 10cm 15cm 15cm,width=1.0\linewidth]{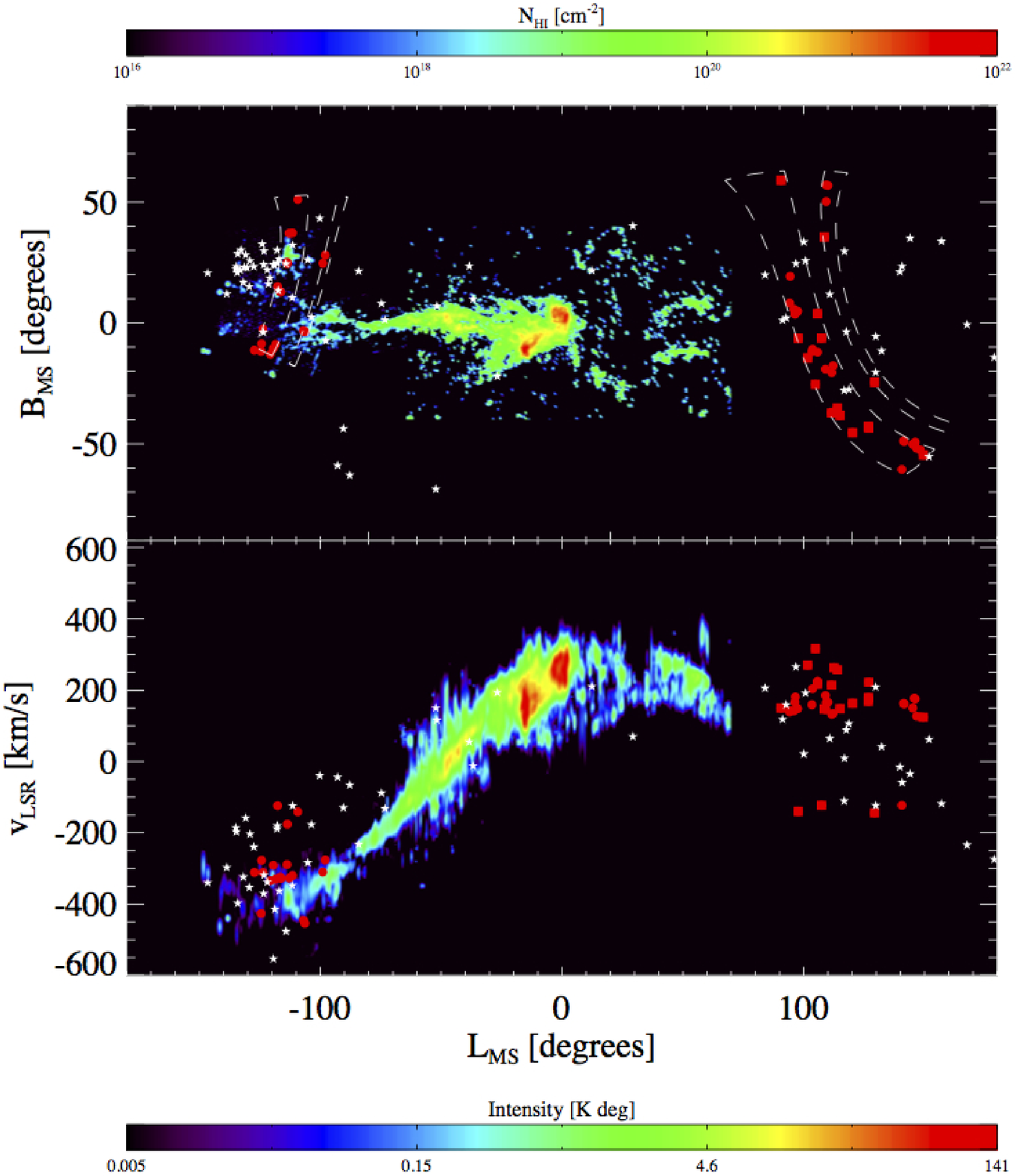}
\caption{The distribution of UCHVCs relative to the MS from N10. 
Coordinates are those of the MS-centric system from \cite{2008ApJ...679..432N}.
The top panel is the spatial distribution of the MS; the x-axis is $L_{MS}$ and
the y-axis $B_{MS}$. 
Shading (color coding in the online edition) of the MS indicates the column density, matching N10.
The \a40 footprint is shown in the top panel by the dashed lines.
The red circles are the UCHVCs,
the most-isolated subsample is indicated by the squares, and the white stars represent LG galaxies. 
The bottom panel is the total intensity of the Magellanic HI integrated along $B_{MS}$ 
(in units of K deg) and shows the kinematics of the MS.
}
\label{fig:ms}
\end{center}
\end{figure}

\begin{figure}
\begin{center}
\includegraphics[keepaspectratio,trim=5cm 1.5cm 4cm 0cm,width=0.9\linewidth]{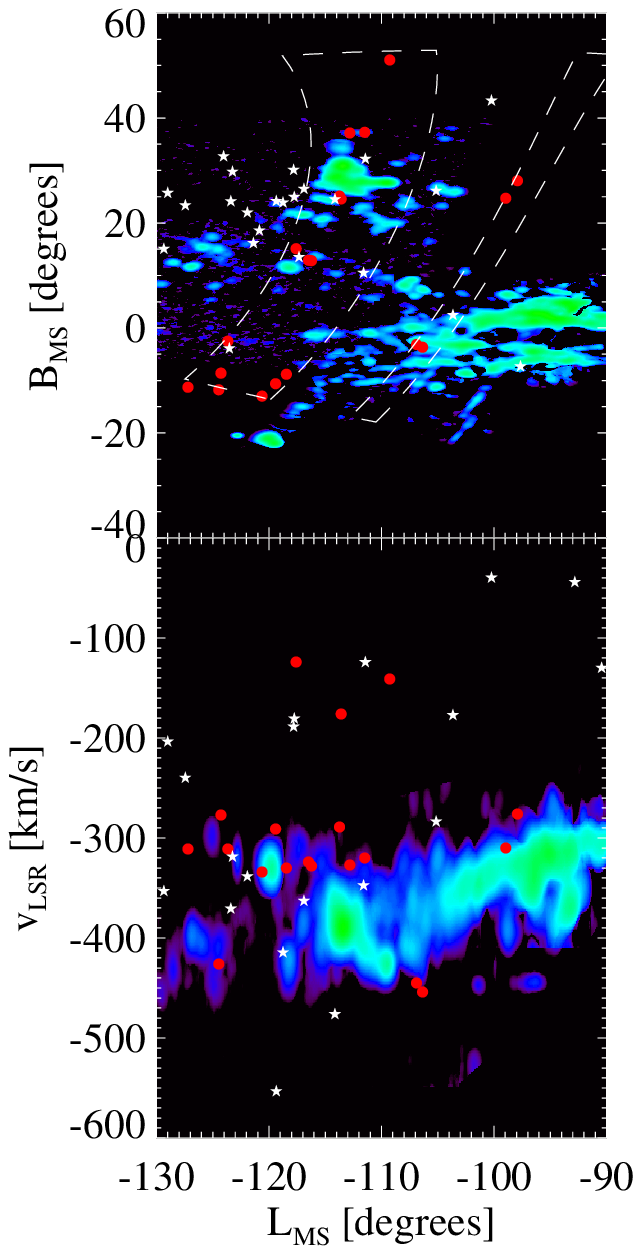}
\caption{A zoomed in view of the fall UCHVCs relative to the MS from \cite{2010ApJ...723.1618N}. Symbols, shading, and panels are
the same as in previous figure.}
\label{fig:ms_zoom}
\end{center}
\end{figure}

For the \a40 footprint,
the fall sky overlaps the tail of the MS and the spring sky is near the known
edge of the LA but not contiguous to it.
N10 extended the known tail of the MS and pointed out its
complexity (see their Figure 4), so we must be especially careful with UCHVCs in the fall sky. 
In Figure \ref{fig:ms}, we show the UCHVCs plotted on the 200\dg\ MS presented in N10. 
The coordinates are the MS-coordinate system of \cite{2008ApJ...679..432N} based on fitting a great circle to the MS,
where L$_{MS}$ is the longitude along the MS and B$_{MS}$ is the latitude above/below the MS. 
The UCHVCs are shown as large symbols (red in the online version) to increase their visibility; 
they are not shown to physical scale nor do their colors match the shading of the MS.
The top panel shows the HI column density of the MS ($\log N_{HI}$ in cm$^{-2}$).
The bottom panel is the total intensity of the MS integrated along $B_{MS}$ (K deg).

In the spring sky, the \a40 footprint approaches but does not overlap the LA of the MS. 
This lack of direct coverage of the MS makes it a challenge to answer the question: 
could the UCHVCs be connected to the LA? 
Future surveys directed at determining  any possible continuation of the LA will 
be able to directly answer this question.
Until then,
the key to answering this question is determining whether the UCHVCs have compatible velocities to be an extension of the LA. 
Clearly, the large velocity spread of UCHVCs seen in the bottom panel of Figure \ref{fig:ms} appears to be incompatible with all of the UCHVCs being associated with the LA. 
Examining models of the MS can provide insight into these questions.
\cite{2006MNRAS.371..108C} model the MS as a tidal structure via interaction with the MW and LMC; they predict that the LA extends to 
L$_{MS}$ $\sim$150\dg\  with a velocity turn over starting from L$_{MS}$$\sim$60\dg\ at \vlsr$\sim$300 \kms\ extending to $\sim$-150 \kms.
In contrast, \cite{2010ApJ...721L..97B} simulate a first passage of the Magellanic Clouds and find a MS that extends to L$_{MS}$$\sim$50\dg\ with a velocity increasing with L$_{MS}$ from \vlsr$\sim$200 to 400 \kms.
If the \cite{2006MNRAS.371..108C} model correctly represents the history of the MS, then the clouds located at \vlsr\ \textless\ 0 \kms\ could be associated with the LA of the MS.
If the \cite{2010ApJ...721L..97B} model is accurate, then the UCHVCs are generally at higher L$_{MS}$ values than predicted by the model but a few of the positive velocity clouds with  $L_{MS}$ \textless\ 100\dg\ and the highest \vlsr\ values may be associated with the MS.
For whichever model of the MS is chosen, some of the UCHVCs could be associated with the LA, but given the large spread in \vlsr\ of the UCHVCs, it is impossible to associate all of the UCHVCs with the LA.

In the fall sky, the \a40 footprint overlaps the extension of the MS detailed in N10. 
In Figure \ref{fig:ms_zoom} we offer a zoomed in view focusing on the fall UCHVCs compared to the MS from N10. 
Here, there clearly appears to be strong overlap between the UCHVCs and the known MS system. 
The three clouds in the fall sky at \vlsr\ \textgreater\ -200 \kms\ appear to be kinematically separated from the MS.
Two other clouds at L$_{MS}$ $\sim$-100\dg\ appear to potentially be spatially separated from the MS but the apparent separation could easily be a result of the coverage of observations of the MS.
However, it is still possible that some of these UCHVCs do indeed represent galaxies. Many of the UCHVCs that overlap with the MS are also in the direction of the M31 subgroup.
Disentangling the gas of known galaxies at a similar velocity from the MS is a long standing problem;
see
\cite{2009ApJ...696..385G} for illustrative examples. 
This is also illustrated in Figure \ref{fig:ms_zoom}, where several LG galaxies are spatially and kinematically coincident with the MS.

\subsection{UCHVCs in the Context of Galactic Halo Clouds}\label{sec:galhalo}
Previous studies have uncovered a population of compact clouds associated with the Galactic halo \citep[e.g.][]{2002ApJ...580L..47L,2005ASPC..331...59L,2006ApJ...637..366S,2006ApJ...653.1210S,2010ApJ...722..367F,2010A&A...509A..60D}.
While well separated from the Galactic hydrogen, these clouds typically have low \vlsr\ values,
and they generally appear to be consistent with Galactic rotation.
The Galactic halo clouds with the most extreme velocities of \cite{2006ApJ...637..366S} have \vlsr\ ranging
from $\lesssim$100 \kms\ to 165 \kms.
The compact halo clouds also tend to be cold clouds, with the vast majority of reported clouds having
$W_{50}$ \textless\ 10 \kms.
Given these characteristics of the halo clouds, the UCHVCs appear as a distinct population.
The UCHVCs appear to universally be warm clouds with linewidths greater than 15 \kms.
In addition, many of the UCHVCs have substantial velocities ($|$\vlsr$| > 200$ \kms) that are difficult to account for in a Galactic halo
model.

\subsection{UCHVCs in the Context of the Small Cloud Population of GALFA-HI}\label{sec:galfa}

GALFA-HI is a survey of neutral hydrogen in the Galaxy 
which, like ALFALFA, uses the ALFA multi beam receiver on the Arecibo 305m antenna.
For GALFA-HI, the IF signal is sent to a different spectrometer than that used
by ALFALFA and is restricted to a $\sim$7 MHz bandpass centered on 1420 MHz.
As a result, the GALFA-HI
 survey has a velocity resolution of 0.184 \kms~ and covers a velocity range of $\pm$700 \kms.
It should be noted that much of the GALFA data is taken commensally with the ALFALFA data through the TOGS program.
Hence comparison of the results of the two surveys provides a check on our signal processing approach.
\citet{2010ApJ...722..395B} presented an initial catalog of compact clouds from the GALFA-HI
survey, and \citet[hereafter S12]{2012ApJ...758...44S} recently released a catalog of compact clouds for the full initial
data release of the GALFA-HI survey.
Herein we focus on the compact clouds of S12 as the most extensive catalog of the compact cloud
population discovered in the GALFA-HI survey and examine how the UCHVCs of this work are related. 

\begin{figure}
\begin{center}
\includegraphics[keepaspectratio,width=0.95\linewidth]{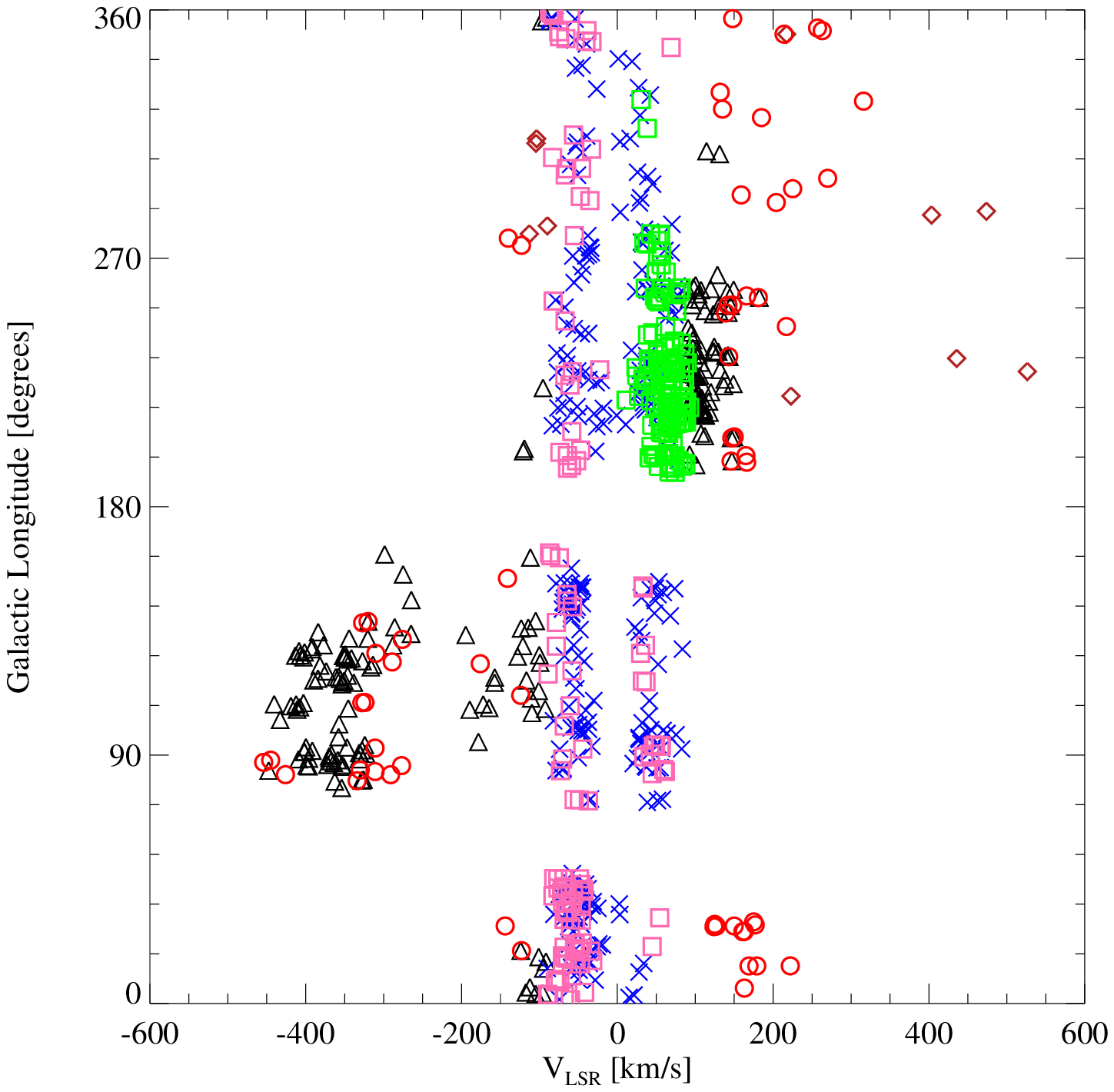}
\caption{
The distribution of UCHVCs in \vlsr$-l$ space compared to the compact cloud populations of GALFA. Symbols and coloring
follow those used in S12: Xs (blue in online version) are the cold low velocity clouds, black squares (pink in the online version) are warm low velocity clouds, grey squares (green in the online version) are the warm low velocity clouds in the third Galactic quadrant, black triangles are the high velocity clouds, and diamonds (dark red in the online version) are the galaxy candidates.
The UCHVCs of this work are shown as circles (bright red in the online version).
}
\label{fig:v_l_galfa}
\end{center}
\end{figure}

The initial major differences to note between the catalog of S12 and the UCHVCs are additional 
selection criteria for the UCHVCs: the limited range of velocities considered and the strong isolation criteria. 
A vast majority of the compact clouds from S12 do not meet these additional criteria.
S12 note several populations of clouds in their catalog which they classify by velocity, linewidth and isolation.
They split between warm and cold clouds at a linewidth of 15 \kms, or a temperature of $\sim$5000 K.
It should be noted that while ALFALFA does not have the velocity resolution of the GALFA-HI survey,
the velocity resolution of $\sim$10 \kms\ is sufficient to distinguish warm from cold clouds;
as can be seen in Figure \ref{fig:meas_hist}, the UCHVCs are all warm clouds with linewidths greater than
15 \kms.
S12 also split their clouds into low velocity and high velocity populations at $|$\vlsr$| = $90 \kms.
They find a few cold clouds with \vlsr\ \textgreater\ 90 \kms, but the vast majority of their cold clouds
are at lower velocities and associated with the Galactic disk, a very distinct population
from the ALFALFA UCHVCs.
The populations from S12 of most relevance to this work are their HVC population ($|$\vlsr$|$ \textgreater\ 90 \kms) and galaxy candidate population; both of these populations are generally composed of warm clouds.
The difference between the HVC population and galaxy candidate population of S12 is that the galaxy candidates 
have an additional stringent isolation criterion (different from the isolation criteria used here) and
hence are the population most directly comparable to the UCHVCs.
In Figure \ref{fig:v_l_galfa}, we compare the distribution of the UCHVCs to the compact clouds of S12 in galactic longitude versus \vlsr. 
In the second Galactic quadrant, the UCHVCs overlap with the HVCs of S12.
This corresponds to the fall sky, and, as noted in the previous section, when considering a stricter isolation
criterion for separation from larger HVC complexes akin to that used by S12, the fall UCHVCs cannot be considered isolated structures. 
In the first and fourth Galactic quadrants, the UCHVCs as a population appear separated from the compact clouds of S12. The positive velocity clouds in the first quadrant and the clouds (at both positive and negative velocities) in the fourth quadrant have no HVC population counterpart in the GALFA compact cloud catalog.
Especially in the fourth quadrant, there are multiple clouds at substantial velocites (\vlsr\ \textgreater\ 200 \kms) that appear well separated from other clouds populations.

\begin{figure}
\begin{center}
\includegraphics[keepaspectratio,trim=1cm 0cm 0cm 0cm,width=0.95\linewidth]{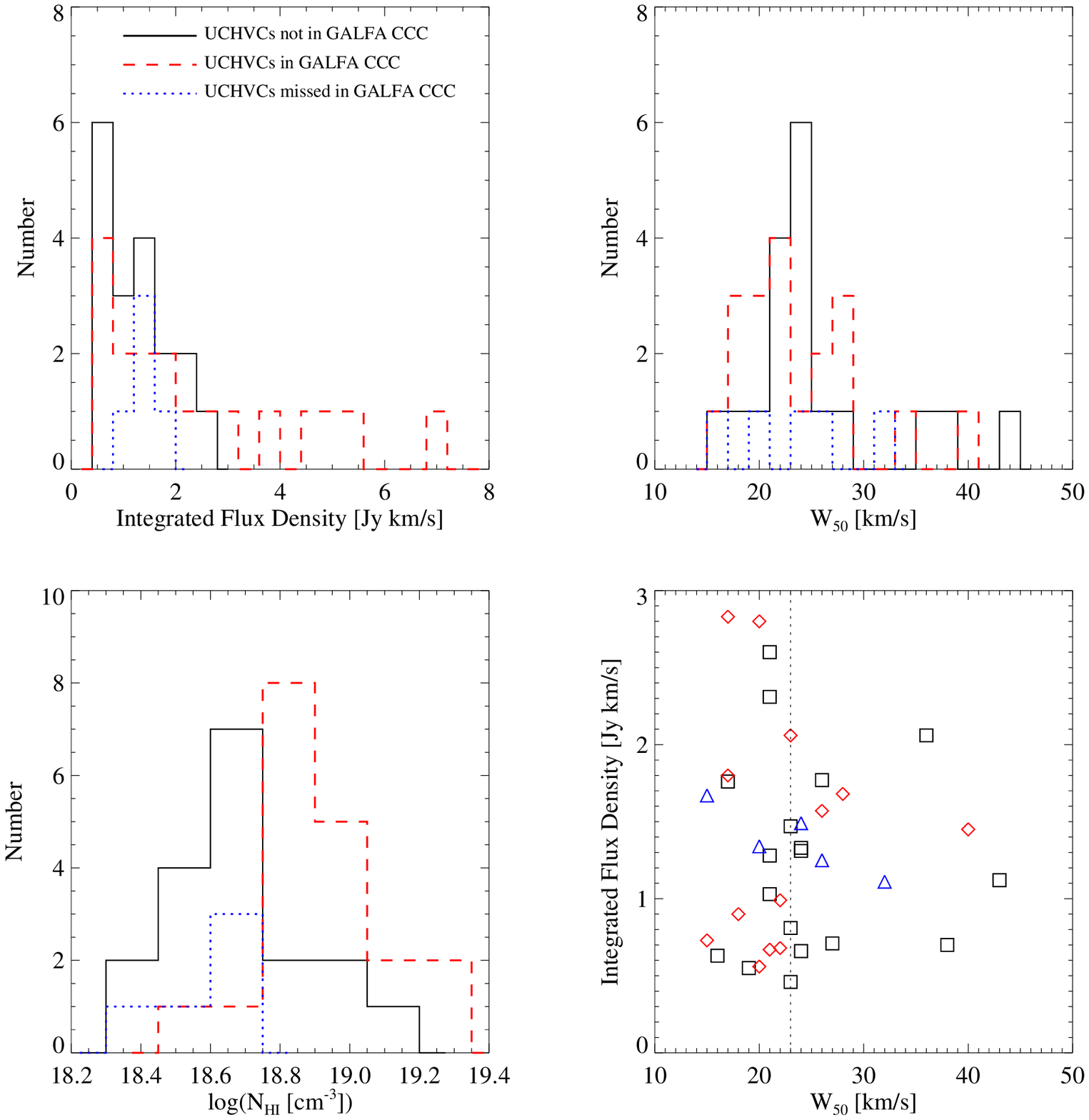}
\caption{
Properties for the UCHVCs not seen in the GALFA-HI dataset (solid lines/squares; black in the online version), 
UCHVCs included in the GALFA compact cloud catalog (CCC) of S12 and those found by the identification algorithm
but discarded from the final catalog (dashed lines/diamonds; red in the online version),
and UCHVCs seen in the GALFA-HI dataset but missed by the cloud finding algorithm of S12 (dotted lines/trianges; blue in the online version).
The dotted line in the bottom right panel indicates the median velocity width of the UCHVCs.
}
\label{fig:compare_galfa}
\end{center}
\end{figure}

As a check of our methodology and dataset, we also perform a direct comparison of the ALFALFA UCHVCs to the catalog of S12.
First, we examine which of the S12 galaxy candidates appear in the \a40 catalog.
S12 find 28 HVCs that they consider extremely isolated and which they classify as galaxy candidates.
Of these, 10 are within the \a40 footprint. 
Two of the GALFA galaxy candidates are classified as extragalactic sources in \a40 (AGC191803 and AGC227874) and are clearly associated
with optical counterparts;
a third S12 galaxy candidate is associated with UGC 7753, a large barred spiral galaxy.
Four of their galaxy candidates are within the ALFALFA data but have $|$\vlsr$|$ \textless\ 120 \kms\ and are not included in this work
(one is included in they \a40 catalog, AGC238801).
One of the galaxy candidates is also included here in the UCHVC catalog -- HVC351.17+58.56+214.
Two of the S12 galaxy candidates are not seen in the ALFALFA data; these are both lower S/N sources (S/N \textless\ 7) and one
is extremely narrow velocity width ($W_{50}=$ 3.9 \kms).

Secondly, we can examine the UCHVCs for counterparts in the S12 catalog.
\ngalfa\ of the \ntot\ UCHVCs are included in the GALFA compact cloud catalog, of which one (HVC351.17+58.56+214) 
is classified by S12 as a galaxy candidate; the other ten are included in their HVC sample.
Seventeen of the UCHVCs are not included in the data coverage of the GALFA DR1 release (D. Saul, private communication); 
these sources
are in the  spring sky region of $\delta=8-16$\dg, where GALFA DR1 has limited coverage because GALFA-HI observations
 started one year after the commencement of ALFALFA data taking and hence commensal data for
that time period are missing.
Of the thirty-one UCHVCs with GALFA coverage not contained within the catalog of S12, eight of these sources
are found by the algorithm but discarded due to either failing the S12 criteria or data quality issues, such as noise spikes.
Five are seen in the data  but not found by the signal identification algorithm of S12. The last eighteen are not visible
in the GALFA-HI data (D. Saul, private communication 2013).
In Figure \ref{fig:compare_galfa}, we explore the differences in properties between the UCHVCs found in the dataset of
GALFA-HI by the signal identification algorithm of the S12 (including sources discarded from the final catalog),
the UCHVCs visible in the GALFA data but not identified by their automated algorithm, and the UCHVCs not visible in the GALFA data.
Most strikingly, there is a bimodal distribution in the average column density with the UCHVCs not visible in the GALFA-HI data 
having the lowest average column densities.
In addition, there is a velocity width effect; generally the UCHVCs identified within the GALFA dataset are the narrowest
velocity width sources. 
In the bottom right panel of Figure \ref{fig:compare_galfa}, we focus on UCHVCs with integrated flux densities less than 3 Jy \kms\ as the higher flux sources are all detected
in the GALFA-HI data.
Then, there are 18 UCHVCs with linewidths greater than 23 \kms, the median $W_{50}$ of the full sample.
Of these, only three are identified in the GALFA-HI dataset and those still tend to be among the highest flux objects with
integrated flux densities greater than 1.45 Jy \kms, above the median value of 1.34 Jy \kms.
The UCHVCs that are identified within the GALFA dataset that have flux densities below the median value of the UCHVC sample
also have linewidths narrower than the median value of the UCHVCs.
This is a straightforward result of the different focus of the two surveys; the GALFA-HI data are designed to detect narrow velocity width HI features
associated with Galactic hydrogen while the ALFALFA dataset is designed to detect extragalactic HI sources with wider linewidths.
While we will address the completeness and reliability of the UCHVC catalog in future work, we note that six UCHVCs not included in the GALFA catalog have 
all been confirmed as real HI signals via confirmation observations with the Arecibo L-Band Wide receiver (Adams et al. in prep).
In addition, the UCHVCs presented here have strict S/N criteria so the likelihood that many of the UCHVCs are false detections is small.
This demonstrates the utility of the ALFALFA dataset, detection algorithm presented here, and the source inspection.



\section{UCHVCs as Minihalo Candidates}\label{sec:mhc}

The mismatch between observations of low mass galaxies and simulations of dark matter halos remains an outstanding question in understanding both the cosmological paradigm and galaxy formation and evolution. Is the $\Lambda$CDM paradigm incorrect? How does star formation and gas accretion proceed in the lowest mass halos? Finding the lowest mass dark matter halos with baryons can help address these question. In this section, we discuss the possibility that the UCHVCs presented in this paper could represent gas-bearing minihalos.
In this context, a minihalo is  dark matter halo below the critical mass of $\sim 10^{10}$ \msun\ where astrophysical
processes begin to strongly affect the baryon content \citep[e.g.][]{2010AdAst2010E..87H,2006MNRAS.371..401H}

\cite{2002ApJS..143..419S} examined in detail how neutral hydrogen could exist in minihalos. 
They found that the neutral gas would be surrounded by an envelope of ionized gas, 
with the specifics depending upon the pressure of the ionized medium the halo is immersed in.
They examined both cuspy (NFW) and constant density (Burkert) cores.
Cuspy cores are predicted by simulations, while observations of dwarf galaxies
indicates that low mass dark matter halos have constant density cores.
The UCHVCs appear to match well the \cite{2002ApJS..143..419S} minihalo models with a median Burkert density profile,  
$D_{HI} \simeq 1.4$ kpc, $M_{HI} \simeq 3 \times 10^5\ M_{\odot}$, total to neutral gas mass ratio of 15, 
peak $N_{HI} \simeq 4 \times 10^{19} \, \mathrm{cm}^{-2}$, total halo mass $M_{vir} \simeq 3 \times 10^8\ M_{\odot}$, 
surrounded by a hot, ionized IGM of pressure $P_{HIM} = 10 \, \mathrm{cm}^{-3}$ K.
The measured column densities  are averaged over the size of the cloud and smeared by the 3\arcmin.5 beam of the Arecibo telescope and hence represent a lower limit to the true peak column density, and so they are consistent with the higher peak $N_{HI}$ values of the model.
The measured $M_{dyn}$ is an estimate of the total mass within the HI extent; 
the total size of the dark matter halo exceeds the HI size by a factor of several, 
explaining the discrepancy between the total halo mass of the model and the inferred dynamical mass from ALFALFA.
Work is ongoing to match the individual UCHVC detections to specific individual models (Y. Faerman et al., submitted).

\subsection{Previous Searches for Minihalos}
A LG origin for HVCs, or at least a subset of the HVC population has been considered before.
With the advent of large-scale, sensitive, blind HI surveys, interest was revived in HVCs as tracers of dark matter halos.
\cite{1999ApJ...514..818B} and   \cite{1999A&A...341..437B}   both postulated a LG origin for HVCs;
\cite{1999A&A...341..437B} specifically proposed that compact HVCs (CHVCs),
identified by their isolation and undisturbed spatial structure, were good candidates to represent dark matter halos throughout the LG.
\cite{2002A&A...391..159D} extracted a set of CHVCs from the Leiden/Dwingeloo Survey (LDS), and \cite{2002AJ....123..873P} similarly presented a set of CHVCs from the HI Parkes All-Sky Survey (HIPASS).
Further work, both observational and theoretical, since the discovery of the CHVC population suggests that they most likely represent a circumgalactic population.
The properties of the CHVC population from the two catalogs are summarized in Table \ref{tab:context}.
Sequentially, this table lists: object class, distance (in kpc), HI angular diameter (in arcmin),
HI diameter (in kpc), peak column density, $W_{50}$, integrated flux density, HI mass, and
dynamical mass within the HI extent.
\cite{2002A&A...392..417D} showed that the properties of the CHVCs for the two datasets are the same when accounting for the better spatial resolution and sensitivity of HIPASS and the better velocity resolution of LDS.

\cite{2002ApJS..143..419S} and  \cite{2003ApJ...589..270M} independently modeled gas in dark matter halos to understand the CHVC population.
Based on considerations of their astrophysical properties,
both groups concluded that the best interpretation of the CHVCs was as circumgalactic objects at $d \lesssim 200$ kpc.
\cite{2002ApJS..143..419S} found that if the CHVCs were at $d > 750$ kpc, their dark matter halos were extremely underconcentrated.
They found that at $d \lesssim 150$ kpc, the CHVCs were consistent with being gas pressure confined in dark matter halos.
In this scenario, the CHVCs represent the subhalos surrounding the Milky Way from its hierarchical formation.
Both pointed out that the gas of the CHVCs must be largely ionized, implying that the total mass of gas is much greater than the observed mass.
If the CHVCs were at distances of 0.7-1 Mpc, extremely low dark-matter-to-gas ratios would then be required to match the observed linewidths of the CHVCs,
and they would violate the $\Lambda$CDM mass-concentration relation.
They argued that the CHVCs must be at $d \lesssim 200$ kpc to match size and total dark matter constraints.
More recent observational evidence also indicates that the CHVCs must be at circumgalactic distances.
The HI masses of the CHVCs at LG distances of $\sim$1 Mpc are a few times $10^7$ \msun, large enough that they should have been detected in surveys of other galaxy groups but have not \citep[e.g.][]{2007ApJ...662..959P,2011AJ....142..137C,2001MNRAS.325.1142Z,2001A&A...375..219B,2004ApJ...610L..17P}.
In addition, higher resolution observations of CHVCs show clear ram pressure indicators in many cases, indicating that the CHVCs are located at circumgalactic distances \citep{2005A&A...432..937W}.
Observations of potential CHVC analogs around M31 also point to a circumgalactic origin.
\cite{2005A&A...436..101W}  studied HVCs associated with M31 in high resolution; importantly, the association of these HVCs with M31 allows a distance constraint to be derived.
As outlined in Table \ref{tab:context}, the properties of the M31 HVCs are a good match to the properties of the CHVCs at $d \sim 150$ kpc, indicating that the two samples are likely a similar population.

\begin{deluxetable*}
{lccccccccl} 
\tablewidth{0pt}
\tabletypesize{\scriptsize}
\tablecaption{HI Content in the LG - HVCs and Galaxies  \label{tab:context}}
\tablehead{
\colhead{Class}  & \colhead{d} & \colhead{$\theta$}    & \colhead{$D_{HI}$} & \colhead{$N_{HI}$} &
\colhead{$W_{50}$} & \colhead{$S_{21}$}  &
\colhead{$M_{HI}$} & \colhead{$M_{tot}$}& \colhead{Refs\tablenotemark{a}} \\
{} & kpc & \arcmin & kpc & atoms $\mathrm{cm}^2$ &  \kms & Jy \kms & \msun & \msun & {}
}
\startdata
UCHVCs       & $d=1000$ & 10 &  2.9$d$    & $\gtrsim 0.6\times 10^{19}$   &  23   &   1.26  &   $1.8\times 10^5 \, d^2 $     & $3.3\times 10^7 \, d$   &     1\\ 
CHVCs (LDS)   &150  &  60  &  2.6   & $1.3\times 10^{19}$    &  25   &    102 &   $5.4\times 10^5$  & $3.5\times 10^7$       &       2\\
CHVCs (HIPASS)&150  &  24  &  0.52 & $1.4\times 10^{19}$    &  35   &    19.9  &  $ 1.1\times 10^5$   & $2.7\times 10^7$       &       3\\
M31 HVCs       &780 &  4.6 &  1.04 & $3.9\times 10^{19}$    &  24 &    2.1  &   $3.0 \times 10^5$  & $4.5\times 10^7$      &       4\\
Leo T          &420 &  5   &  0.6  & $70\times 10^{19}$     &  16     &   6.7   &   $2.8 \times 10^5 $       & $.33\times 10^7$    &  5\\
Leo P         & $1750$  & 2.0 & 1.0  & $20\times 10^{19}$   &  24         &  1.31      &  $9.5\times 10^5$  &     $1.3\times 10^7$ & 6,7,8\\         
\hline
\enddata
\tablenotetext{a}
{References: 1: this work, 2: \cite{2002A&A...391..159D}, 3: \cite{2002AJ....123..873P}, 
4:\cite{2005A&A...436..101W},  5: \cite{2008MNRAS.384..535R},
6: \cite{LeoP_HI}, 7: \cite{LeoP_optical}, 8:\cite{LeoP_metallicity}
}
\end{deluxetable*}

Multiple searches have been undertaken for minihalos around nearby galaxy groups
\citep[e.g.][]{2001MNRAS.325.1142Z,2001A&A...375..219B,2002A&A...382...43D,2003MNRAS.346..787M,2004MNRAS.351..333B,2004ApJ...610L..17P,2007ApJ...662..959P,2011ApJS..197...28P,2009AJ....138..287C,2009MNRAS.400..743K,2009ApJ...692.1447I,2011AJ....142..137C,2011AJ....141....9C,2012ApJ...761..186M}.
Generally, these surveys must choose between sensitivity and coverage area.
\cite{2009ApJ...692.1447I} undertook a deep survey of the nearby isolated galaxy NGC 2903 sensitive to an HI mass of $2 \times 10^5$ \msun\ and covering 150 kpc $\times$ 260 kpc.
This survey was sensitive enough to (barely) detect a Leo T analog but given that the survey footprint only extends to $\sim$100 kpc in projected radius from the galaxy center, detection of an object at $\gtrsim$400 kpc from the galaxy center would depend strongly on orientation. \cite{2009ApJ...692.1447I} did detect one minihalo with an HI mass of $2.6 \times 10^6$ \msun, a comparable stellar stellar mass and a dynamical mass of $\gtrsim 10^8$ \msun.
\cite{2011AJ....141....9C} undertook a large (480 kpc $\times$ 1.2 Mpc; 8\dg.7 $\times$ 21\dg.3) survey centered on the region between the M81/M82 and NGC 2403 galaxy groups.
Their survey had a mass detection limit of $3.2 \times 10^6$ \msun which is not deep enough to detect a Leo T analog. 
While their survey covers a large footprint, it is focused on the region between two connected galaxy groups and coverage of the outskirts of the galaxy groups is limited.
They detect several massive HI clouds (M \textgreater\ $10^6$ \msun) and determine that these clouds likely arise from tidal processes given their clustering near M81.
\cite{2012ApJ...761..186M} surveyed the M101 group over 1050 $\times$ 825 kpc (8\dg.5 $\times$ 6\dg.7) to a mass senstivity of varying from 2 to 10 $\times 10^6$ \msun\ over their footprint.
This footprint includes all objects out to $\sim$400 kpc from the central galaxy, regardless of orientation, but the survey is not sensitive enough to detect a Leo T analog.
They do identify a new low surface brightness dwarf galaxy through an HI detection and a starless HI cloud with an HI mass of $1.2 \times 10^7$ \msun.

\subsection{Known Minihalos in the LG}
In considering the UCHVCs as gas-bearing minihalos in the LG, we first want to examine the context of the LG
and ask what we may empirically expect a minihalo to look like.
The population of the LG has increased substantially in the last few years with the discovery of the UFD
satellites of the Milky Way from automated stellar searches of the Sloan Digital Sky Survey \citep{2010AdAst2010E..21W} 
and targeted searches for satellites of M31 \citep[e.g.][]{2007ApJ...671.1591I,2009Natur.461...66M}.
The UFDs have indicative dynamical masses within the baryon extent of $10^6 - 10^7$ \msun\ and most likely inhabit dark matter halos that
qualify them as minihalos.
With the exception of Leo T 
and the recently discovered Leo P,
the UFDs are located within the virial radius of the MW or M31 and have no detectable gas content.

Surveys of low mass galaxies in the field indicate that, with large scatter, dwarf galaxies tend to be gas-rich
and can have atomic gas as their dominant baryon component \citep[e.g.][]{2006ApJ...653..240G,2001AJ....121.2420S}.
Modulo the uncertainties in how astrophysical processes affect the baryon content of the lowest mass halos,
one would naively expect the trend of high gas fraction to continue as lower mass galaxies are discovered.
Leo T is the only UFD discovered through optical surveys that has neutral gas content; 
it is also the UFD that is most distant from the MW.
The other UFDs are located within the virial radius of the MW or M31 and 
many show signs of tidal interaction with the MW 
\citep[e.g.][]{2012ApJ...756...79S}.
\cite{2009ApJ...696..385G} find that morphological segregation is strong in the LG with dwarf galaxies within 270 kpc
of the Milky Way or Andromeda showing no evidence of neutral gas content.
Leo T is on the edge of detectability for SDSS; were it located further away, its stellar population would not have been detected \citep{2010AdAst2010E...8K}.
Taken together, these facts raise the possibility that more gas-rich UFDs are lurking in the LG with distances and stellar populations that would leave them undetected in SDSS.

\begin{figure}
\begin{center}
\includegraphics[keepaspectratio,width=0.9\linewidth]{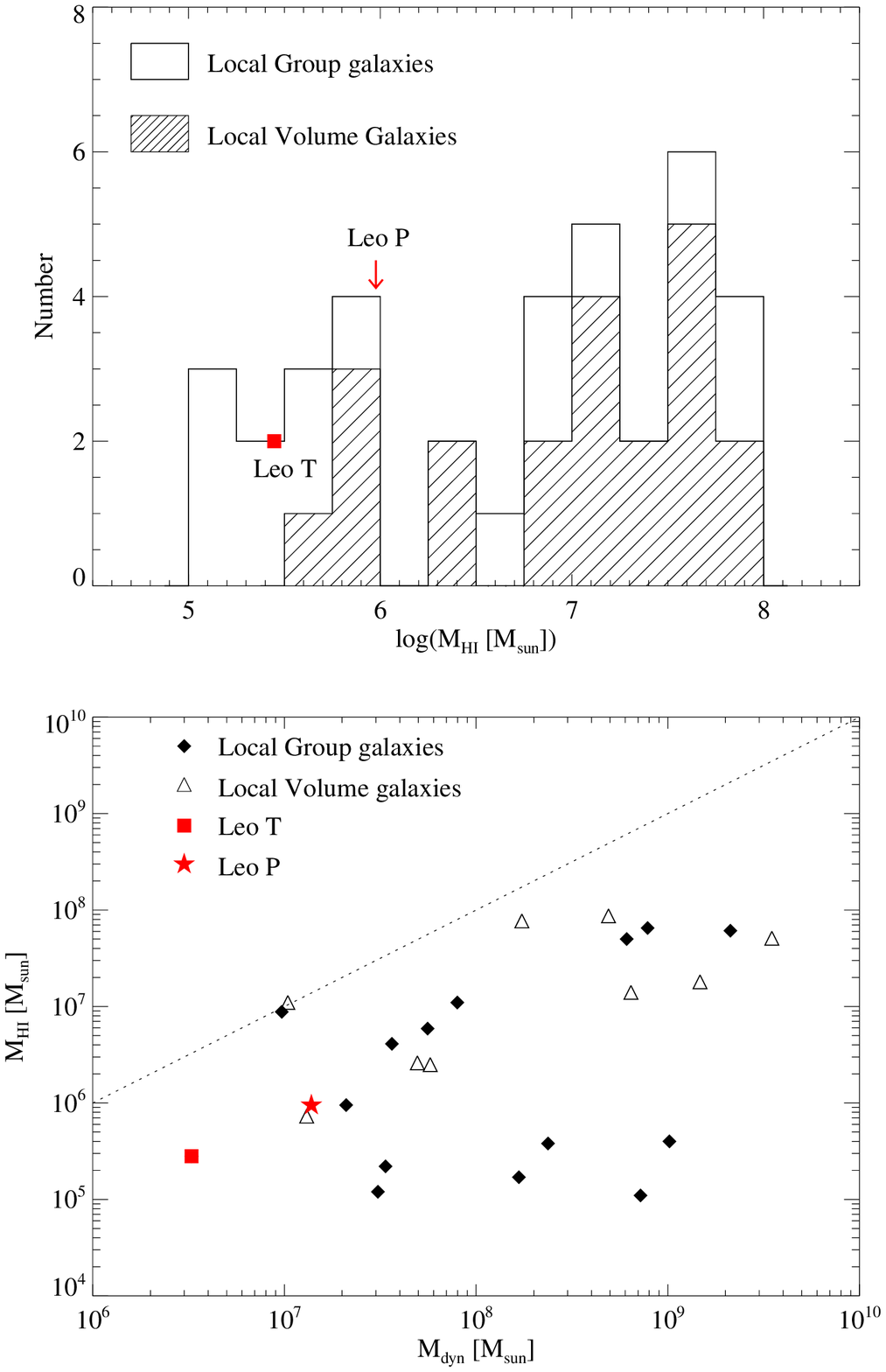}
\caption{The top panel is a histogram of HI mass in the LG and nearby dwarf galaxies in the Local Volume (indicated by the hashed histogram),
 including Leo T
(its contribution indicated by the red filled square), from the catalog of \cite{2012AJ....144....4M}. 
The location of Leo P is also indicated. 
The bottom panel is HI mass as a function of dynamical mass within the baryon extent. The diamonds are LG
galaxies with HI content, the triangles Local Volume dwarfs, the filled square is Leo T and the filled star is Leo P. 
The dynamical masses are compiled from the literature and are calculated using a variety of different methods and
at different extents of the galaxies; in all
cases the dynamical masses are underestimates of the true dynamical mass
\citep{2009MNRAS.394L.102L,2006MNRAS.369.1321D,2010ApJ...711..361G,1989A&A...214...33S,1999PASP..111..306C,1996ApJS..105..269H,1998ARA&A..36..435M,2008MNRAS.384..535R,2007AJ....133.2242K,2004A&A...413..525B,2012ApJ...751...46K,1988A&A...198...33S,2005MNRAS.359L..53B,2006MNRAS.365.1220B}.
The dotted line indicates where M$_{dyn}$ equals M$_{HI}$.
In addition to having low HI masses, Leo T and Leo P also have low dynamical masses.
}
\label{fig:hilg}
\end{center}
\end{figure}

Leo T serves as our prototype of what a gas-rich minihalo will look like;
it has motivated our search for more minihalos and the discovery of Leo P.
In Figure \ref{fig:hilg} we examine the HI properties of the LG galaxies and neighboring dwarf galaxies within
3 Mpc in comparison to Leo T and Leo P to infer what we may expect for future minihalo detections.
The top panel of 
Figure \ref{fig:hilg} shows a histogram of the HI masses of dwarf galaxies within the LG and neighboring systems, 
taken from \cite{2012AJ....144....4M}.
Leo P and Leo T  have some of the lowest HI masses in the LG and Local Volume (LV);
we would expect previously undetected systems to have low HI masses.
The bottom panel of Figure \ref{fig:hilg} illustrates the parameter space occupied by Leo T and Leo P in the LG and LV;
they have low HI masses \emph{and} low dynamical masses.

\subsection{Evidence for the UCHVCs as Minihalo Candidates}

In assessing the UCHVCs as minihalo candidates,
we first consider if their astrophysical properties are consistent with the scenario.
As mentioned above, the UCHVCs are a good match to the models of \cite{2002ApJS..143..419S}.
Importantly, the UCHVCs also overcome the objections that ruled out the CHVCs as minihalo
candidates throughout the LG.
As summarized in Table \ref{tab:context}, the UCHVCs have HI masses typical of $\sim 10^5 \, d^2$ \msun\ 
and HI diameters of $\sim2.9 \, d$ kpc.
These smaller sizes and lower fluxes suggest that at distances of 1 Mpc, 
the physical properties of the UCHVCs are good matches to
the CHVC properties at distances of $\sim$250 kpc. 
In this scenario, the CHVCs could represent subhalos within the MW and the
UCHVCs represent isolated structures within the LG.

\begin{figure}
\begin{center}
\includegraphics[keepaspectratio,width=0.9\linewidth]{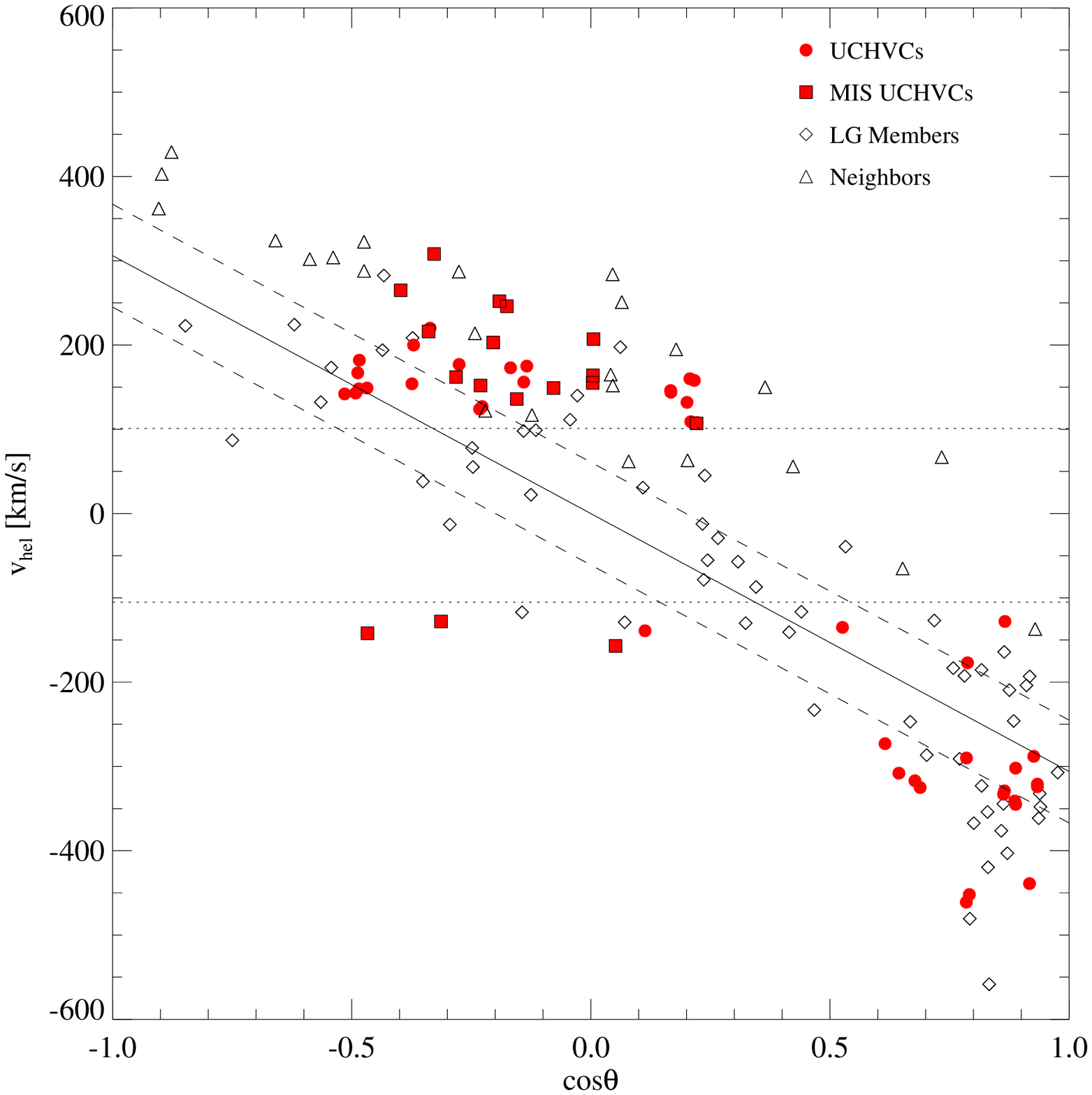}
\caption{Cosine of the angle from the Solar apex versus heliocentric velocity.
The solid line shows the relation of \cite{1999AJ....118..337C} and the dashed
lines are their stated error.
The dotted lines indicate inaccessible velocity space due to the UCHVC velocity selection criterion.
The filled circles (red in the online version) are the UCHVCs with the outlined filled squares (red in the online version)
indicating the MIS UCHVCs.
The  diamonds are the LG galaxies from \cite{2012AJ....144....4M}
and the triangles are neighboring galaxies within 3 Mpc that are not bound to
the LG.
}
\label{fig:costh}
\end{center}
\end{figure}

The LG is a bound group of galaxies, hence studying the kinematics of the UCHVCs
can help constrain their association with the LG.
In Figure \ref{fig:costh} we compare the motions of the UCHVCs to the LG.
Following \cite{1999AJ....118..337C}, we plot \vsun\ versus the cosine of the angle from the LG apex.
In general the UCHVCs show similar behavior to the motions of the LG galaxies,
lending credence to the possibility that they trace LG dark matter halos. 
They do appear to have a
higher velocity dispersion, similar to the nearby neighbor galaxies that are not bound to the LG.
This may suggest that the UCHVCs are outlying systems, marginally bound to the LG.

Finally,  we offer a preliminary comparison of the UCHVCs to the Via Lactea II (VL) simulation of \cite{2008Natur.454..735D},
a high resolution cosmological N-body simulation of a Milky Way analog.
We compare the spatial and kinematic distribution of the UCHVCs to the dark matter halos of the VL simulation to see if the hypothesis of UCHVCS
as minihalos is consistent with theoretical predictions.
We utilize the full volume of the simulation, which includes 20,048 halos that extend to more than 3 Mpc from the central MW analog halo.
In addition to the central massive halo,
there is a second massive halo which
is a fortuitous analog to M31 \citep{2012MNRAS.426.1808T}.
In our favored model, we place this second massive halo at the approximate location of M31 in order
to most closely match the LG.
We also use the original simulation coordinates plus five random orientations of the subhalos
to demonstrate the importance of structure within the LG.
After the coordinate transformations, we only consider the halos within the simulation that lie within the boundaries
of the \a40 coverage and meet our velocity criterion.

\begin{figure}
\begin{center}
\includegraphics[keepaspectratio,trim=1cm 0cm 0cm 0cm,width=0.95\linewidth]{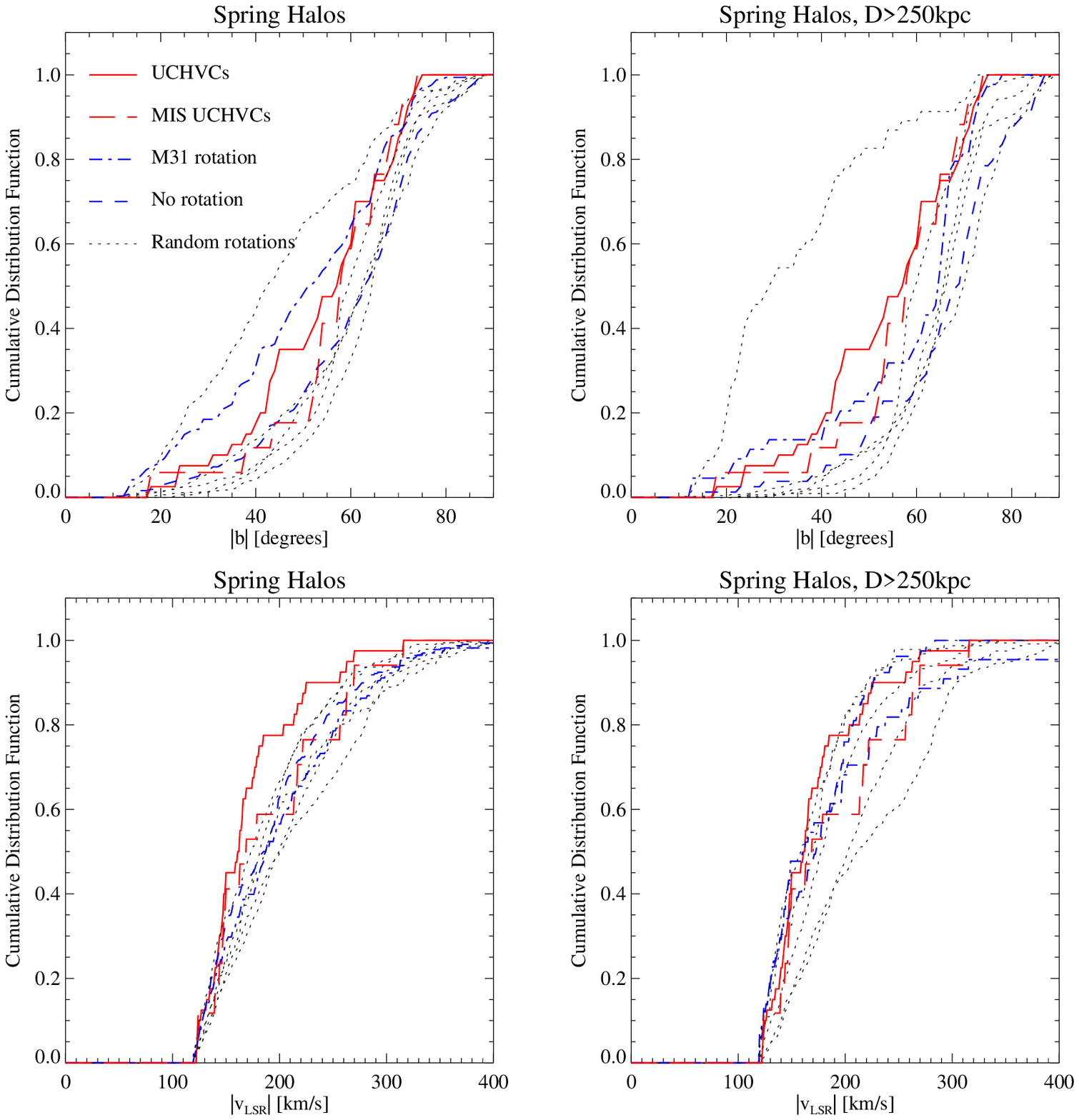}
\caption{
The distribution of subhalos from the Via Lactea II simulation compared to the UCHVCs (solid line, red in the online version)
 and the most-isolated subsample (dashed line, red in the online version). 
The dot-dash line (blue in the online version) represents the subhalos in the original simulation coordinate system; 
the dashed line (blue in the online version)
is our favored orientation where
the simulation rotated to place the second massive halo at the approximate location of M31. 
The dotted lines represent five random rotations of the simulation coordinates. 
The left-hand column shows the distribution of all the VL subhalos in the spring footprint that meet our velocity criterion,
and the right-hand column shows the VL subhalos that are located further than 250 kpc from the central massive halo.
Overall, the UCHVCs appear consistent with the distribution of halos from the simulation,
especially for our favored orientation.  
Given the large differences between halo distribution depending on the rotation of the simulation
coordinates, it is clear that accounting for structure is crucial.
}
\label{fig:vl}
\end{center}
\end{figure}

In Figure \ref{fig:vl} we show the distribution of galactic latitude  and \vlsr\ for the UCHVCs and the VL subhalos. 
Due to the presence of large and complex HVC structure in the fall sky, we focus on the spring sky for our comparison.
In the left column we show all the halos that match our selection criteria;
 in the right column we show only those halos located further than 250 kpc from the central massive halo
 to more closely approximate the halos we expect to be gas-bearing.
The effects of structure are much more noticeable when only the most distant halos are considered; 
the different orientations show a much wider spread in the distribution of $|b|$ in this case.
The galactic latitude plot is especially important as it provides a quick test of whether the distribution of 
clouds is within the Galactic disk or a circumgalactic distribution.
If the UCHVCs are associated with the Galactic disk, a flattened distribution of $|b|$ values is expected compared to
the case if the UCHVCs are distributed around the Galaxy. 
The UCHVCs and MIS UCHVCs have similar distributions for $|b|$ and $|$\vlsr$|$.
The favored orientation of the VL simulation appears to match well the distribution
of $|b|$ for the UCHVCs.
The large differences in the cumulative distribution function (CDF) of $|b|$ for the
random orientations shows the importance of structure.
The kinematics of the UCHVCs appear to be consistent with the VL simulation in all cases
with the CDF of $|$\vlsr$|$ matching well in all cases.
While it is beyond the scope of this paper to do a full halo-population analysis, the rough
analysis presented here shows that the UCHVCs agree reasonably well with the VL simulation.

We can also use the VL simulations to provide a rough check of the numbers of halos expected.
There are \nspring\ UCHVCs in the spring sky, including \nmh\ in the most-isolated subsample.
We compare to our favored orientation of the VL simulation, 
noting that  it matches the spring sky in that we are looking into the outskirts of 
the simulation as the spring region of ALFALFA probes the outskirts of the LG.
There are a total of 168 VL halos that meet our velocity criterion
in the region of the simulation that matches the \a40 spring footprint.
When limited to halos with distances from the central MW analog halo greater than 
250 kpc, there are a total of 44 halos; 27 of these halos
have $M_{tidal} > 10^7$ \msun. 
Given the roughness of our numbers the two populations appear to be consistent.

\subsection{The UCHVCs as Galaxies}
As galaxies, the UCHVCs would favor the outskirts of the LG, 
rather than the central regions, with distances of $\sim$500 kpc -- 1 Mpc.
They would have HI masses of $\sim 10^{5}$ \msun\ with envelopes of warm ionized hydrogen with masses of $\sim 10^6$ \msun.
The indicative dynamical masses within the HI extent are $\sim 10^7 - 10^8$ \msun, and the total hosting halo masses
are likely $\gtrsim 10^9$ \msun.
While this hypothesis is attractive, it cannot be definitively proven until distance constraints
are in place for the UCHVCs.
Further work is necessary in order to constrain their distances
as the ALFALFA HI detection carries no direct distance information.
The kinematics of the UCHVCs are dominated by LG interactions,
 so the velocity cannot offer any insights to the distance.
The detection of an optical counterpart can constrain the distance through
studies of the stellar population.
It is also possible to constrain the distance solely through HI by
using synthesis imaging to determine the rotational
velocity of the UCHVCs and constrain the distance
through the baryonic Tully-Fisher relation
\citep[e.g.][]{LeoP_HI,2012AJ....143...40M}.
An alternative to confirming the distance of the UCHVCs directly is to detect UCHVC analogs around
other nearby galaxy groups and use the association with the group to constrain the distance and properties of the clouds.
Planned future HI surveys using phased-array-feeds will be able to robustly detect these objects.

Confirming that a subset of the UCHVCs are galaxies will 
offer many insights.
The UCHVCs will
increase the number of low-mass galaxies known in the Local Volume,
decreasing the discrepancy between simulations and observations.
In addition, the UCHVCs will
trace the outskirts of the LG allowing the comparison between simulations and observations to be extended
to a larger volume.
The UCHVCs will also serve as isolated examples of the lowest mass galaxies, having not yet interacted substantially with the MW.
The UCHVCs offer the potential to study star formation in extreme, low metallicity environments
as the presence of gas
means there is a possibility of star formation.
In fact, Leo T has recently formed stars and Leo P has ongoing star formation with one HII region.
Abundance measurements of the HII region in Leo P indicate that it is among the lowest metallicity systems known 
and blind HI surveys may prove to be a promising way to detect low luminosity,
extremely metal deficient galaxies \citep{LeoP_metallicity}.

The two confirmed low mass gas-rich galaxies in the Local Volume, Leo T and Leo P, both have high average column densities
and small HI angular diameters,
as can be seen in Figures \ref{fig:meas_hist} and \ref{fig:inf_hist}.
It may be reasonable to expect then that the most compact and highest column density UCHVCs are the best candidates to
represent low-mass gas-rich galaxies.
HVC274.68+74.70-123,  HVC351.17+58.56+214, and HVC13.59+54.52+169 are in the most-isolated subsample,
have average angular diameters \textless\ 7\arcmin, and
have $\bar N_{HI} > 10^{19}\, \mathrm{cm}^{-2}$;
we suggest that these are the best galaxy candidates in our sample. 
One of these candidates, HVC351.17+58.56+214 is also identified by the GALFA-HI survey as a good galaxy candidate.
Notably, it is among the most compact clouds included in this catalog (7\arcmin\ $\times$ 5\arcmin) and has one of the highest
column densities ($\log~\bar N_{HI} = 19.3$).
If we adopt a representative distance of 1 Mpc, it has a HI mass of $3.9 \times 10^5$ \msun\ and an indicative
dynamical mass within the HI extent of $2.1 \times 10^7$\msun.


\section{Conclusion}\label{sec:conclusion}

We present a set of \ntot\ ultra-compact high velocity clouds
which are of interest as speculative minihalo candidates.
In brief, the properties of the UCHVCs are summarized below.
\begin{itemize}
\item They have HI integrated flux densities from 0.66--8.55 Jy \kms\ with a median of 
\fmedian\ Jy \kms, 
linewidths of 15--70 \kms\ with a median of \wmedian\ \kms, and
angular diameters of 4--20\arcmin\ with a median of \hsizemedian \arcmin.
\item They are selected according to strict isolation criteria.
As a result, they are distinct from known HVC populations.
\item Their HI sizes and HI fluxes  allow them to overcome previous objections
leveled against CHVCs as LG minihalos.
\item They are consistent with the minihalo models of \cite{2002ApJS..143..419S}.
At a distance of $\sim$1 Mpc, they have HI masses of $10^5 - 10^6$ \msun\ and
dynamical masses within the HI extent of $10^7 - 10^8$ \msun.
Their total gas masses, including the surrounding ionized envelope,
would be $\sim 10^6-10^7$\msun\ and the total hosting halo masses would be
$\lesssim 10^9$ \msun.
\item As galaxies, they would allow us to probe the outskirts of the LG, 
study low mass systems that have remained isolated from the MW, and provide an avenue
for indentifying extremely metal deficient galaxies.
\end{itemize}

We thank David Nidever for providing his dataset of the Magellanic Stream, Bart Wakker for
providing the updated HVC catalog and useful comments, Destry Saul for help 
in comparing the UCHVCs to the GALFA-HI dataset, and the anonymous referee for
helpful comments.
The ALFALFA survey team at Cornell is supported by NSF grants AST-0607007 and 
AST-1107390 to RG and MPH and by grants from the Brinson Foundation. 
EAKA has also been supported by an NSF predoctoral fellowship.

This research has made use of NASA's Astrophysics Data System Bibliographic Services and the NASA/IPAC Extragalactic Database (NED), 
which is operated by the Jet Propulsion Laboratory, California Institute of Technology, under contract with the Nataion Aeronautics
 and Space Administration.


\bibliography{refs}

\begin{thebibliography}{97}
\expandafter\ifx\csname natexlab\endcsname\relax\def\natexlab#1{#1}\fi

\bibitem[{{Barnes} \& {de Blok}(2004)}]{2004MNRAS.351..333B}
{Barnes}, D.~G., \& {de Blok}, W.~J.~G. 2004, \mnras, 351, 333

\bibitem[{{Begum} \& {Chengalur}(2004)}]{2004A&A...413..525B}
{Begum}, A., \& {Chengalur}, J.~N. 2004, \aap, 413, 525

\bibitem[{{Begum} {et~al.}(2006){Begum}, {Chengalur}, {Karachentsev}, {Kaisin},
  \& {Sharina}}]{2006MNRAS.365.1220B}
{Begum}, A., {Chengalur}, J.~N., {Karachentsev}, I.~D., {Kaisin}, S.~S., \&
  {Sharina}, M.~E. 2006, \mnras, 365, 1220

\bibitem[{{Begum} {et~al.}(2005){Begum}, {Chengalur}, {Karachentsev}, \&
  {Sharina}}]{2005MNRAS.359L..53B}
{Begum}, A., {Chengalur}, J.~N., {Karachentsev}, I.~D., \& {Sharina}, M.~E.
  2005, \mnras, 359, L53

\bibitem[{{Begum} {et~al.}(2010){Begum}, {Stanimirovi{\'c}}, {Peek},
  {Ballering}, {Heiles}, {Douglas}, {Putman}, {Gibson}, {Grcevich}, {Korpela},
  {Lee}, {Saul}, \& {Gallagher}}]{2010ApJ...722..395B}
{Begum}, A., {Stanimirovi{\'c}}, S., {Peek}, J.~E., {et~al.} 2010, \apj, 722,
  395

\bibitem[{{Besla} {et~al.}(2010){Besla}, {Kallivayalil}, {Hernquist}, {van der
  Marel}, {Cox}, \& {Kere{\v s}}}]{2010ApJ...721L..97B}
{Besla}, G., {Kallivayalil}, N., {Hernquist}, L., {et~al.} 2010, \apjl, 721,
  L97

\bibitem[{{Blanton} {et~al.}(2005){Blanton}, {Lupton}, {Schlegel}, {Strauss},
  {Brinkmann}, {Fukugita}, \& {Loveday}}]{2005ApJ...631..208B}
{Blanton}, M.~R., {Lupton}, R.~H., {Schlegel}, D.~J., {et~al.} 2005, \apj, 631,
  208

\bibitem[{{Blitz} {et~al.}(1999){Blitz}, {Spergel}, {Teuben}, {Hartmann}, \&
  {Burton}}]{1999ApJ...514..818B}
{Blitz}, L., {Spergel}, D.~N., {Teuben}, P.~J., {Hartmann}, D., \& {Burton},
  W.~B. 1999, \apj, 514, 818

\bibitem[{{Bovill} \& {Ricotti}(2011)}]{BR11}
{Bovill}, M.~S., \& {Ricotti}, M. 2011, \apj, 741, 17

\bibitem[{{Braun} \& {Burton}(1999)}]{1999A&A...341..437B}
{Braun}, R., \& {Burton}, W.~B. 1999, \aap, 341, 437

\bibitem[{{Braun} \& {Burton}(2001)}]{2001A&A...375..219B}
---. 2001, \aap, 375, 219

\bibitem[{{Cannon} {et~al.}(2011){Cannon}, {Giovanelli}, {Haynes},
  {Janowiecki}, {Parker}, {Salzer}, {Adams}, {Engstrom}, {Huang}, {McQuinn},
  {Ott}, {Saintonge}, {Skillman}, {Allan}, {Erny}, {Fliss}, \&
  {Smith}}]{2011ApJ...739L..22C}
{Cannon}, J.~M., {Giovanelli}, R., {Haynes}, M.~P., {et~al.} 2011, \apjl, 739,
  L22

\bibitem[{{Chynoweth} {et~al.}(2011{\natexlab{a}}){Chynoweth},
  {Holley-Bockelmann}, {Polisensky}, \& {Langston}}]{2011AJ....142..137C}
{Chynoweth}, K.~M., {Holley-Bockelmann}, K., {Polisensky}, E., \& {Langston},
  G.~I. 2011{\natexlab{a}}, \aj, 142, 137

\bibitem[{{Chynoweth} {et~al.}(2011{\natexlab{b}}){Chynoweth}, {Langston}, \&
  {Holley-Bockelmann}}]{2011AJ....141....9C}
{Chynoweth}, K.~M., {Langston}, G.~I., \& {Holley-Bockelmann}, K.
  2011{\natexlab{b}}, \aj, 141, 9

\bibitem[{{Chynoweth} {et~al.}(2009){Chynoweth}, {Langston},
  {Holley-Bockelmann}, \& {Lockman}}]{2009AJ....138..287C}
{Chynoweth}, K.~M., {Langston}, G.~I., {Holley-Bockelmann}, K., \& {Lockman},
  F.~J. 2009, \aj, 138, 287

\bibitem[{{Connors} {et~al.}(2006){Connors}, {Kawata}, \&
  {Gibson}}]{2006MNRAS.371..108C}
{Connors}, T.~W., {Kawata}, D., \& {Gibson}, B.~K. 2006, \mnras, 371, 108

\bibitem[{{Cook} {et~al.}(1999){Cook}, {Mateo}, {Olszewski}, {Vogt}, {Stubbs},
  \& {Diercks}}]{1999PASP..111..306C}
{Cook}, K.~H., {Mateo}, M., {Olszewski}, E.~W., {et~al.} 1999, \pasp, 111, 306

\bibitem[{{Courteau} \& {van den Bergh}(1999)}]{1999AJ....118..337C}
{Courteau}, S., \& {van den Bergh}, S. 1999, \aj, 118, 337

\bibitem[{{de Blok} {et~al.}(2002){de Blok}, {Zwaan}, {Dijkstra}, {Briggs}, \&
  {Freeman}}]{2002A&A...382...43D}
{de Blok}, W.~J.~G., {Zwaan}, M.~A., {Dijkstra}, M., {Briggs}, F.~H., \&
  {Freeman}, K.~C. 2002, \aap, 382, 43

\bibitem[{{de Heij} {et~al.}(2002{\natexlab{a}}){de Heij}, {Braun}, \&
  {Burton}}]{2002A&A...392..417D}
{de Heij}, V., {Braun}, R., \& {Burton}, W.~B. 2002{\natexlab{a}}, \aap, 392,
  417

\bibitem[{{de Heij} {et~al.}(2002{\natexlab{b}}){de Heij}, {Braun}, \&
  {Burton}}]{2002A&A...391..159D}
---. 2002{\natexlab{b}}, \aap, 391, 159

\bibitem[{{De Rijcke} {et~al.}(2006){De Rijcke}, {Prugniel}, {Simien}, \&
  {Dejonghe}}]{2006MNRAS.369.1321D}
{De Rijcke}, S., {Prugniel}, P., {Simien}, F., \& {Dejonghe}, H. 2006, \mnras,
  369, 1321

\bibitem[{{Dedes} \& {Kalberla}(2010)}]{2010A&A...509A..60D}
{Dedes}, L., \& {Kalberla}, P.~W.~M. 2010, \aap, 509, A60

\bibitem[{{Diemand} {et~al.}(2008){Diemand}, {Kuhlen}, {Madau}, {Zemp},
  {Moore}, {Potter}, \& {Stadel}}]{2008Natur.454..735D}
{Diemand}, J., {Kuhlen}, M., {Madau}, P., {et~al.} 2008, \nat, 454, 735

\bibitem[{{Dieter}(1965)}]{1965AJ.....70..552D}
{Dieter}, N.~H. 1965, \aj, 70, 552

\bibitem[{{Evoli} {et~al.}(2011){Evoli}, {Salucci}, {Lapi}, \&
  {Danese}}]{2011ApJ...743...45E}
{Evoli}, C., {Salucci}, P., {Lapi}, A., \& {Danese}, L. 2011, \apj, 743, 45

\bibitem[{{Ford} {et~al.}(2010){Ford}, {Lockman}, \&
  {McClure-Griffiths}}]{2010ApJ...722..367F}
{Ford}, H.~A., {Lockman}, F.~J., \& {McClure-Griffiths}, N.~M. 2010, \apj, 722,
  367

\bibitem[{{Fukugita} \& {Peebles}(2006)}]{2006ApJ...639..590F}
{Fukugita}, M., \& {Peebles}, P.~J.~E. 2006, \apj, 639, 590

\bibitem[{{Geha} {et~al.}(2006){Geha}, {Blanton}, {Masjedi}, \&
  {West}}]{2006ApJ...653..240G}
{Geha}, M., {Blanton}, M.~R., {Masjedi}, M., \& {West}, A.~A. 2006, \apj, 653,
  240

\bibitem[{{Geha} {et~al.}(2010){Geha}, {van der Marel}, {Guhathakurta},
  {Gilbert}, {Kalirai}, \& {Kirby}}]{2010ApJ...711..361G}
{Geha}, M., {van der Marel}, R.~P., {Guhathakurta}, P., {et~al.} 2010, \apj,
  711, 361

\bibitem[{{Giovanelli} {et~al.}(2013){Giovanelli}, {Haynes}, {Adams}, {Cannon},
  {Rhode}, {Salzer}, {Skillman}, \& {McQuinn}}]{LeoP_HI}
{Giovanelli}, R., {Haynes}, M.~P., {Adams}, E.~A.~K., {et~al.} 2013, submitted

\bibitem[{{Giovanelli} {et~al.}(2010){Giovanelli}, {Haynes}, {Kent}, \&
  {Adams}}]{2010ApJ...708L..22G}
{Giovanelli}, R., {Haynes}, M.~P., {Kent}, B.~R., \& {Adams}, E.~A.~K. 2010,
  \apjl, 708, L22

\bibitem[{{Giovanelli} {et~al.}(2005){Giovanelli}, {Haynes}, {Kent},
  {Perillat}, {Saintonge}, {Brosch}, {Catinella}, {Hoffman}, {Stierwalt},
  {Spekkens}, {Lerner}, {Masters}, {Momjian}, {Rosenberg}, {Springob},
  {Boselli}, {Charmandaris}, {Darling}, {Davies}, {Garcia Lambas}, {Gavazzi},
  {Giovanardi}, {Hardy}, {Hunt}, {Iovino}, {Karachentsev}, {Karachentseva},
  {Koopmann}, {Marinoni}, {Minchin}, {Muller}, {Putman}, {Pantoja}, {Salzer},
  {Scodeggio}, {Skillman}, {Solanes}, {Valotto}, {van Driel}, \& {van
  Zee}}]{2005AJ....130.2598G}
{Giovanelli}, R., {Haynes}, M.~P., {Kent}, B.~R., {et~al.} 2005, \aj, 130, 2598

\bibitem[{{Giovanelli} {et~al.}(2007){Giovanelli}, {Haynes}, {Kent},
  {Saintonge}, {Stierwalt}, {Altaf}, {Balonek}, {Brosch}, {Brown}, {Catinella},
  {Furniss}, {Goldstein}, {Hoffman}, {Koopmann}, {Kornreich}, {Mahmood},
  {Martin}, {Masters}, {Mitschang}, {Momjian}, {Nair}, {Rosenberg}, \&
  {Walsh}}]{2007AJ....133.2569G}
---. 2007, \aj, 133, 2569

\bibitem[{{Grcevich} \& {Putman}(2009)}]{2009ApJ...696..385G}
{Grcevich}, J., \& {Putman}, M.~E. 2009, \apj, 696, 385

\bibitem[{{Guo} {et~al.}(2010){Guo}, {White}, {Li}, \&
  {Boylan-Kolchin}}]{2010MNRAS.404.1111G}
{Guo}, Q., {White}, S., {Li}, C., \& {Boylan-Kolchin}, M. 2010, \mnras, 404,
  1111

\bibitem[{{Hartmann} \& {Burton}(1997)}]{1997agnh.book.....H}
{Hartmann}, D., \& {Burton}, W.~B. 1997,

\bibitem[{{Haynes} {et~al.}(2011){Haynes}, {Giovanelli}, {Martin}, {Hess},
  {Saintonge}, {Adams}, {Hallenbeck}, {Hoffman}, {Huang}, {Kent}, {Koopmann},
  {Papastergis}, {Stierwalt}, {Balonek}, {Craig}, {Higdon}, {Kornreich},
  {Miller}, {O'Donoghue}, {Olowin}, {Rosenberg}, {Spekkens}, {Troischt}, \&
  {Wilcots}}]{2011AJ....142..170H}
{Haynes}, M.~P., {Giovanelli}, R., {Martin}, A.~M., {et~al.} 2011, \aj, 142,
  170

\bibitem[{{Hoeft} \& {Gottl{\"o}ber}(2010)}]{2010AdAst2010E..87H}
{Hoeft}, M., \& {Gottl{\"o}ber}, S. 2010, Advances in Astronomy, 2010, 87

\bibitem[{{Hoeft} {et~al.}(2006){Hoeft}, {Yepes}, {Gottl{\"o}ber}, \&
  {Springel}}]{2006MNRAS.371..401H}
{Hoeft}, M., {Yepes}, G., {Gottl{\"o}ber}, S., \& {Springel}, V. 2006, \mnras,
  371, 401

\bibitem[{{Hoffman} {et~al.}(1996){Hoffman}, {Salpeter}, {Farhat}, {Roos},
  {Williams}, \& {Helou}}]{1996ApJS..105..269H}
{Hoffman}, G.~L., {Salpeter}, E.~E., {Farhat}, B., {et~al.} 1996, \apjs, 105,
  269

\bibitem[{{Ibata} {et~al.}(2007){Ibata}, {Martin}, {Irwin}, {Chapman},
  {Ferguson}, {Lewis}, \& {McConnachie}}]{2007ApJ...671.1591I}
{Ibata}, R., {Martin}, N.~F., {Irwin}, M., {et~al.} 2007, \apj, 671, 1591

\bibitem[{{Irwin} {et~al.}(2009){Irwin}, {Hoffman}, {Spekkens}, {Haynes},
  {Giovanelli}, {Linder}, {Catinella}, {Momjian}, {Koribalski}, {Davies},
  {Brinks}, {de Blok}, {Putman}, \& {van Driel}}]{2009ApJ...692.1447I}
{Irwin}, J.~A., {Hoffman}, G.~L., {Spekkens}, K., {et~al.} 2009, \apj, 692,
  1447

\bibitem[{{Karachentsev} \& {Makarov}(1996)}]{1996AJ....111..794K}
{Karachentsev}, I.~D., \& {Makarov}, D.~A. 1996, \aj, 111, 794

\bibitem[{{Kepley} {et~al.}(2007){Kepley}, {Wilcots}, {Hunter}, \&
  {Nordgren}}]{2007AJ....133.2242K}
{Kepley}, A.~A., {Wilcots}, E.~M., {Hunter}, D.~A., \& {Nordgren}, T. 2007,
  \aj, 133, 2242

\bibitem[{{Kirby} {et~al.}(2012){Kirby}, {Cohen}, \&
  {Bellazzini}}]{2012ApJ...751...46K}
{Kirby}, E.~N., {Cohen}, J.~G., \& {Bellazzini}, M. 2012, \apj, 751, 46

\bibitem[{{Klypin} {et~al.}(1999){Klypin}, {Kravtsov}, {Valenzuela}, \&
  {Prada}}]{1999ApJ...522...82K}
{Klypin}, A., {Kravtsov}, A.~V., {Valenzuela}, O., \& {Prada}, F. 1999, \apj,
  522, 82

\bibitem[{{Kova{\v c}} {et~al.}(2009){Kova{\v c}}, {Oosterloo}, \& {van der
  Hulst}}]{2009MNRAS.400..743K}
{Kova{\v c}}, K., {Oosterloo}, T.~A., \& {van der Hulst}, J.~M. 2009, \mnras,
  400, 743

\bibitem[{{Kravtsov}(2010)}]{2010AdAst2010E...8K}
{Kravtsov}, A. 2010, Advances in Astronomy, 2010, 8

\bibitem[{{Lockman}(2002)}]{2002ApJ...580L..47L}
{Lockman}, F.~J. 2002, \apjl, 580, L47

\bibitem[{{Lockman} \& {Pidopryhora}(2005)}]{2005ASPC..331...59L}
{Lockman}, F.~J., \& {Pidopryhora}, Y. 2005, in Astronomical Society of the
  Pacific Conference Series, Vol. 331, Extra-Planar Gas, ed. R.~{Braun}, 59

\bibitem[{{{\L}okas}(2009)}]{2009MNRAS.394L.102L}
{{\L}okas}, E.~L. 2009, \mnras, 394, L102

\bibitem[{{Maloney} \& {Putman}(2003)}]{2003ApJ...589..270M}
{Maloney}, P.~R., \& {Putman}, M.~E. 2003, \apj, 589, 270

\bibitem[{{Martin} {et~al.}(2010){Martin}, {Papastergis}, {Giovanelli},
  {Haynes}, {Springob}, \& {Stierwalt}}]{2010ApJ...723.1359M}
{Martin}, A.~M., {Papastergis}, E., {Giovanelli}, R., {et~al.} 2010, \apj, 723,
  1359

\bibitem[{{Martin} {et~al.}(2008){Martin}, {de Jong}, \&
  {Rix}}]{2008ApJ...684.1075M}
{Martin}, N.~F., {de Jong}, J.~T.~A., \& {Rix}, H.-W. 2008, \apj, 684, 1075

\bibitem[{{Mateo}(1998)}]{1998ARA&A..36..435M}
{Mateo}, M.~L. 1998, \araa, 36, 435

\bibitem[{{Mathewson} {et~al.}(1974){Mathewson}, {Cleary}, \&
  {Murray}}]{1974ApJ...190..291M}
{Mathewson}, D.~S., {Cleary}, M.~N., \& {Murray}, J.~D. 1974, \apj, 190, 291

\bibitem[{{McConnachie}(2012)}]{2012AJ....144....4M}
{McConnachie}, A.~W. 2012, \aj, 144, 4

\bibitem[{{McConnachie} {et~al.}(2009){McConnachie}, {Irwin}, {Ibata},
  {Dubinski}, {Widrow}, {Martin}, {C{\^o}t{\'e}}, {Dotter}, {Navarro},
  {Ferguson}, {Puzia}, {Lewis}, {Babul}, {Barmby}, {Bienaym{\'e}}, {Chapman},
  {Cockcroft}, {Collins}, {Fardal}, {Harris}, {Huxor}, {Mackey},
  {Pe{\~n}arrubia}, {Rich}, {Richer}, {Siebert}, {Tanvir}, {Valls-Gabaud}, \&
  {Venn}}]{2009Natur.461...66M}
{McConnachie}, A.~W., {Irwin}, M.~J., {Ibata}, R.~A., {et~al.} 2009, \nat, 461,
  66

\bibitem[{{McGaugh}(2012)}]{2012AJ....143...40M}
{McGaugh}, S.~S. 2012, \aj, 143, 40

\bibitem[{{Meyer} {et~al.}(2004){Meyer}, {Zwaan}, {Webster}, {Staveley-Smith},
  {Ryan-Weber}, {Drinkwater}, {Barnes}, {Howlett}, {Kilborn}, {Stevens},
  {Waugh}, {Pierce}, {Bhathal}, {de Blok}, {Disney}, {Ekers}, {Freeman},
  {Garcia}, {Gibson}, {Harnett}, {Henning}, {Jerjen}, {Kesteven}, {Knezek},
  {Koribalski}, {Mader}, {Marquarding}, {Minchin}, {O'Brien}, {Oosterloo},
  {Price}, {Putman}, {Ryder}, {Sadler}, {Stewart}, {Stootman}, \&
  {Wright}}]{2004MNRAS.350.1195M}
{Meyer}, M.~J., {Zwaan}, M.~A., {Webster}, R.~L., {et~al.} 2004, \mnras, 350,
  1195

\bibitem[{{Mihos} {et~al.}(2012){Mihos}, {Keating}, {Holley-Bockelmann},
  {Pisano}, \& {Kassim}}]{2012ApJ...761..186M}
{Mihos}, J.~C., {Keating}, K.~M., {Holley-Bockelmann}, K., {Pisano}, D.~J., \&
  {Kassim}, N.~E. 2012, \apj, 761, 186

\bibitem[{{Minchin} {et~al.}(2003){Minchin}, {Disney}, {Boyce}, {de Blok},
  {Parker}, {Banks}, {Freeman}, {Garcia}, {Gibson}, {Grossi}, {Haynes},
  {Knezek}, {Lang}, {Malin}, {Price}, {Stewart}, \&
  {Wright}}]{2003MNRAS.346..787M}
{Minchin}, R.~F., {Disney}, M.~J., {Boyce}, P.~J., {et~al.} 2003, \mnras, 346,
  787

\bibitem[{{Moore} {et~al.}(1999){Moore}, {Ghigna}, {Governato}, {Lake},
  {Quinn}, {Stadel}, \& {Tozzi}}]{1999ApJ...524L..19M}
{Moore}, B., {Ghigna}, S., {Governato}, F., {et~al.} 1999, \apjl, 524, L19

\bibitem[{{Mu{\~n}oz} {et~al.}(2010){Mu{\~n}oz}, {Geha}, \&
  {Willman}}]{2010AJ....140..138M}
{Mu{\~n}oz}, R.~R., {Geha}, M., \& {Willman}, B. 2010, \aj, 140, 138

\bibitem[{{Nidever} {et~al.}(2008){Nidever}, {Majewski}, \&
  {Burton}}]{2008ApJ...679..432N}
{Nidever}, D.~L., {Majewski}, S.~R., \& {Burton}, W.~B. 2008, \apj, 679, 432

\bibitem[{{Nidever} {et~al.}(2010){Nidever}, {Majewski}, {Butler Burton}, \&
  {Nigra}}]{2010ApJ...723.1618N}
{Nidever}, D.~L., {Majewski}, S.~R., {Butler Burton}, W., \& {Nigra}, L. 2010,
  \apj, 723, 1618

\bibitem[{{Papastergis} {et~al.}(2012){Papastergis}, {Cattaneo}, {Huang},
  {Giovanelli}, \& {Haynes}}]{2012ApJ...759..138P}
{Papastergis}, E., {Cattaneo}, A., {Huang}, S., {Giovanelli}, R., \& {Haynes},
  M.~P. 2012, \apj, 759, 138

\bibitem[{{Papastergis} {et~al.}(2011){Papastergis}, {Martin}, {Giovanelli}, \&
  {Haynes}}]{2011ApJ...739...38P}
{Papastergis}, E., {Martin}, A.~M., {Giovanelli}, R., \& {Haynes}, M.~P. 2011,
  \apj, 739, 38

\bibitem[{{Peek} {et~al.}(2008){Peek}, {Putman}, \&
  {Sommer-Larsen}}]{2008ApJ...674..227P}
{Peek}, J.~E.~G., {Putman}, M.~E., \& {Sommer-Larsen}, J. 2008, \apj, 674, 227

\bibitem[{{Peek} {et~al.}(2011){Peek}, {Heiles}, {Douglas}, {Lee}, {Grcevich},
  {Stanimirovi{\'c}}, {Putman}, {Korpela}, {Gibson}, {Begum}, {Saul},
  {Robishaw}, \& {Kr{\v c}o}}]{2011ApJS..194...20P}
{Peek}, J.~E.~G., {Heiles}, C., {Douglas}, K.~A., {et~al.} 2011, \apjs, 194, 20

\bibitem[{{Pisano} {et~al.}(2004){Pisano}, {Barnes}, {Gibson},
  {Staveley-Smith}, {Freeman}, \& {Kilborn}}]{2004ApJ...610L..17P}
{Pisano}, D.~J., {Barnes}, D.~G., {Gibson}, B.~K., {et~al.} 2004, \apjl, 610,
  L17

\bibitem[{{Pisano} {et~al.}(2007){Pisano}, {Barnes}, {Gibson},
  {Staveley-Smith}, {Freeman}, \& {Kilborn}}]{2007ApJ...662..959P}
---. 2007, \apj, 662, 959

\bibitem[{{Pisano} {et~al.}(2011){Pisano}, {Barnes}, {Staveley-Smith},
  {Gibson}, {Kilborn}, \& {Freeman}}]{2011ApJS..197...28P}
{Pisano}, D.~J., {Barnes}, D.~G., {Staveley-Smith}, L., {et~al.} 2011, \apjs,
  197, 28

\bibitem[{{Putman} {et~al.}(2002){Putman}, {de Heij}, {Staveley-Smith},
  {Braun}, {Freeman}, {Gibson}, {Burton}, {Barnes}, {Banks}, {Bhathal}, {de
  Blok}, {Boyce}, {Disney}, {Drinkwater}, {Ekers}, {Henning}, {Jerjen},
  {Kilborn}, {Knezek}, {Koribalski}, {Malin}, {Marquarding}, {Minchin},
  {Mould}, {Oosterloo}, {Price}, {Ryder}, {Sadler}, {Stewart}, {Stootman},
  {Webster}, \& {Wright}}]{2002AJ....123..873P}
{Putman}, M.~E., {de Heij}, V., {Staveley-Smith}, L., {et~al.} 2002, \aj, 123,
  873

\bibitem[{{Reyes} {et~al.}(2012){Reyes}, {Mandelbaum}, {Gunn}, {Nakajima},
  {Seljak}, \& {Hirata}}]{2012MNRAS.425.2610R}
{Reyes}, R., {Mandelbaum}, R., {Gunn}, J.~E., {et~al.} 2012, \mnras, 425, 2610

\bibitem[{{Rhode} {et~al.}(2013){Rhode}, {Salzer}, {Haurberg}, {Van Sistine},
  {Young}, {Haynes}, {Giovanelli}, {Cannon}, {Skillman}, {McQuinn}, \&
  {Adams}}]{LeoP_optical}
{Rhode}, K.~L., {Salzer}, J.~J., {Haurberg}, N.~C., {et~al.} 2013, accepted

\bibitem[{{Rocha} {et~al.}(2012){Rocha}, {Peter}, \&
  {Bullock}}]{2012MNRAS.425..231R}
{Rocha}, M., {Peter}, A.~H.~G., \& {Bullock}, J. 2012, \mnras, 425, 231

\bibitem[{{Ryan-Weber} {et~al.}(2008){Ryan-Weber}, {Begum}, {Oosterloo}, {Pal},
  {Irwin}, {Belokurov}, {Evans}, \& {Zucker}}]{2008MNRAS.384..535R}
{Ryan-Weber}, E.~V., {Begum}, A., {Oosterloo}, T., {et~al.} 2008, \mnras, 384,
  535

\bibitem[{{Saintonge}(2007)}]{2007AJ....133.2087S}
{Saintonge}, A. 2007, \aj, 133, 2087

\bibitem[{{Sand} {et~al.}(2012){Sand}, {Strader}, {Willman}, {Zaritsky},
  {McLeod}, {Caldwell}, {Seth}, \& {Olszewski}}]{2012ApJ...756...79S}
{Sand}, D.~J., {Strader}, J., {Willman}, B., {et~al.} 2012, \apj, 756, 79

\bibitem[{{Saul} {et~al.}(2012){Saul}, {Peek}, {Grcevich}, {Putman}, {Douglas},
  {Korpela}, {Stanimirovi{\'c}}, {Heiles}, {Gibson}, {Lee}, {Begum}, {Brown},
  {Burkhart}, {Hamden}, {Pingel}, \& {Tonnesen}}]{2012ApJ...758...44S}
{Saul}, D.~R., {Peek}, J.~E.~G., {Grcevich}, J., {et~al.} 2012, \apj, 758, 44

\bibitem[{{Schombert} {et~al.}(2001){Schombert}, {McGaugh}, \&
  {Eder}}]{2001AJ....121.2420S}
{Schombert}, J.~M., {McGaugh}, S.~S., \& {Eder}, J.~A. 2001, \aj, 121, 2420

\bibitem[{{Shostak} \& {Skillman}(1989)}]{1989A&A...214...33S}
{Shostak}, G.~S., \& {Skillman}, E.~D. 1989, \aap, 214, 33

\bibitem[{{Simon} \& {Geha}(2007)}]{2007ApJ...670..313S}
{Simon}, J.~D., \& {Geha}, M. 2007, \apj, 670, 313

\bibitem[{{Skillman} {et~al.}(1988){Skillman}, {Terlevich}, {Teuben}, \& {van
  Woerden}}]{1988A&A...198...33S}
{Skillman}, E.~D., {Terlevich}, R., {Teuben}, P.~J., \& {van Woerden}, H. 1988,
  \aap, 198, 33

\bibitem[{{Skillman} {et~al.}(2013){Skillman}, {Salzer}, {Berg}, {Pogge},
  {Haurberg}, {Cannon}, {Aver}, {Olive}, {Giovanelli}, {Haynes}, {Adams},
  {McQuinn}, \& {Rhode}}]{LeoP_metallicity}
{Skillman}, E.~D., {Salzer}, J.~J., {Berg}, D.~A., {et~al.} 2013, submitted

\bibitem[{{Stanimirovi{\'c}} {et~al.}(2006){Stanimirovi{\'c}}, {Putman},
  {Heiles}, {Peek}, {Goldsmith}, {Koo}, {Kr{\v c}o}, {Lee}, {Mock}, {Muller},
  {Pandian}, {Parsons}, {Tang}, \& {Werthimer}}]{2006ApJ...653.1210S}
{Stanimirovi{\'c}}, S., {Putman}, M., {Heiles}, C., {et~al.} 2006, \apj, 653,
  1210

\bibitem[{{Sternberg} {et~al.}(2002){Sternberg}, {McKee}, \&
  {Wolfire}}]{2002ApJS..143..419S}
{Sternberg}, A., {McKee}, C.~F., \& {Wolfire}, M.~G. 2002, \apjs, 143, 419

\bibitem[{{Stil} {et~al.}(2006){Stil}, {Lockman}, {Taylor}, {Dickey}, {Kavars},
  {Martin}, {Rothwell}, {Boothroyd}, \&
  {McClure-Griffiths}}]{2006ApJ...637..366S}
{Stil}, J.~M., {Lockman}, F.~J., {Taylor}, A.~R., {et~al.} 2006, \apj, 637, 366

\bibitem[{{Strigari} {et~al.}(2008){Strigari}, {Bullock}, {Kaplinghat},
  {Simon}, {Geha}, {Willman}, \& {Walker}}]{2008Natur.454.1096S}
{Strigari}, L.~E., {Bullock}, J.~S., {Kaplinghat}, M., {et~al.} 2008, \nat,
  454, 1096

\bibitem[{{Teyssier} {et~al.}(2012){Teyssier}, {Johnston}, \&
  {Kuhlen}}]{2012MNRAS.426.1808T}
{Teyssier}, M., {Johnston}, K.~V., \& {Kuhlen}, M. 2012, \mnras, 426, 1808

\bibitem[{{Wakker} \& {van Woerden}(1991)}]{1991A&A...250..509W}
{Wakker}, B.~P., \& {van Woerden}, H. 1991, \aap, 250, 509

\bibitem[{{Westmeier} {et~al.}(2005{\natexlab{a}}){Westmeier}, {Braun}, \&
  {Thilker}}]{2005A&A...436..101W}
{Westmeier}, T., {Braun}, R., \& {Thilker}, D. 2005{\natexlab{a}}, \aap, 436,
  101

\bibitem[{{Westmeier} {et~al.}(2005{\natexlab{b}}){Westmeier}, {Br{\"u}ns}, \&
  {Kerp}}]{2005A&A...432..937W}
{Westmeier}, T., {Br{\"u}ns}, C., \& {Kerp}, J. 2005{\natexlab{b}}, \aap, 432,
  937

\bibitem[{{Willman}(2010)}]{2010AdAst2010E..21W}
{Willman}, B. 2010, Advances in Astronomy, 2010, 21

\bibitem[{{Zwaan}(2001)}]{2001MNRAS.325.1142Z}
{Zwaan}, M.~A. 2001, \mnras, 325, 1142

\end{thebibliography}



\end{document}